\documentclass{ieeeaccess}
\usepackage{cite}
\usepackage{amsmath,amssymb,amsfonts}
\usepackage{graphicx}
\usepackage{textcomp}
\usepackage{times}
\usepackage{epsfig}
\usepackage{amsmath}
\usepackage{amssymb}
\usepackage{subfig}
\usepackage{stackengine}
\usepackage{booktabs}
\usepackage{multirow}
\usepackage{url}

\def\BibTeX{{\rm B\kern-.05em{\sc i\kern-.025em b}\kern-.08em
    T\kern-.1667em\lower.7ex\hbox{E}\kern-.125emX}}
\begin{document}
\history{Date of publication June 16, 2020, date of current version June 29, 2020.}
\doi{10.1109/ACCESS.2020.3002921}

\title{Joint Spatial and Angular Super-Resolution from a Single Image}
\author{\uppercase{Andre Ivan}\authorrefmark{1}, \IEEEmembership{Member, IEEE},
\uppercase{Williem}\authorrefmark{2}, \IEEEmembership{Member, IEEE}, and \uppercase{In Kyu park}\authorrefmark{1},
\IEEEmembership{Senior Member, IEEE}}
\address[1]{Department of Information and Communication Engineering, Inha University, Incheon 22212, Korea (e-mail: andreivan13@gmail.com, pik@inha.ac.kr)}
\address[2]{Verihubs, Indonesia (e-mail: williem@verihubs.com)}
\tfootnote{This work was supported in part by the Samsung Research Funding Center of Samsung Electronics under Project SRFC-IT1702-06, in part by the Institute of Information and Communications Technology Planning and Evaluation (IITP) funded by the Korea Government (MSIT) (Development of Acceleration SW Platform Technology for On-device Intelligent Information Processing in Smart Devices) under Grant 2017-0-00142, and in part by the Artificial Intelligence Convergence Research Center (Inha University) under Grant 2020-0-01389.}

\markboth
{A. Ivan \headeretal: Joint Spatial and Angular Super-Resolution from a Single Image}
{A. Ivan \headeretal: Joint Spatial and Angular Super-Resolution from a Single Image}

\corresp{Corresponding author: In Kyu Park (e-mail: pik@inha.ac.kr).}

\begin{abstract}
Synthesizing a densely sampled light field from a single image is highly beneficial for many applications. Moreover, jointly solving both angular and spatial super-resolution problem also introduces new possibilities in light field imaging. The conventional method relies on physical-based rendering and a secondary network to solve the angular super-resolution problem. In addition, pixel-based loss limits the network capability to infer scene geometry globally. In this paper, we show that both super-resolution problems can be solved jointly from a single image by proposing a single end-to-end deep neural network that does not require a physical-based approach. Two novel loss functions based on known light field domain knowledge are proposed to enable the network to consider the relation between sub-aperture images. Experimental results show that the proposed model successfully synthesizes dense high resolution light field and it outperforms the state-of-the-art method in both quantitative and qualitative criteria. The method can be generalized to various scenes, rather than focusing on a particular subject. The synthesized light field can be used as if it has been captured by a light field camera, such as depth estimation and refocusing.
\end{abstract}

\begin{keywords}
Deep neural network, Light field, Machine learning, Super-resolution
\end{keywords}

\titlepgskip=-15pt

\maketitle

\section{Introduction}
Light fields have attracted considerable interest from computer vision and graphic communities due to their capability to capture multiple light rays from various directions.
Recent studies utilized densely sampled light field captured by off-the-shelf light field cameras.
Many applications, such as depth estimation~\cite{Schilling_CVPR18,Shin_CVPR18,Williem_CVPR16,Williem_PAMI17}, refocusing~\cite{Ng_CSTR05}, and 3D reconstruction~\cite{Johannsen_ACCV16,Vianello_CVPR18}, exploit the rich information of a light field image.
\begin{figure}[t]
\begin{center}
	{\includegraphics[width=0.16\linewidth]{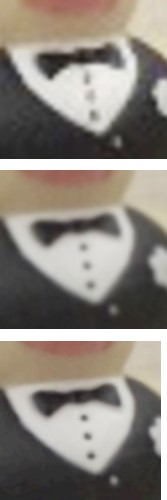}}~%
  	{\includegraphics[width=0.383\linewidth]{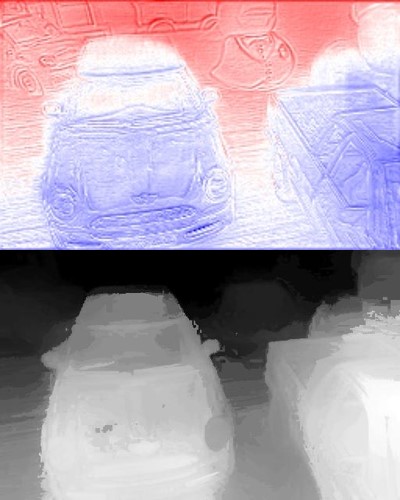}}~%
	{\includegraphics[width=0.383\linewidth]{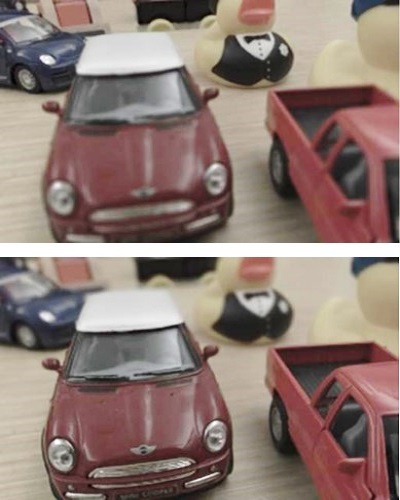}}~%
	
	\subfloat[]{\includegraphics[width=0.16\linewidth]{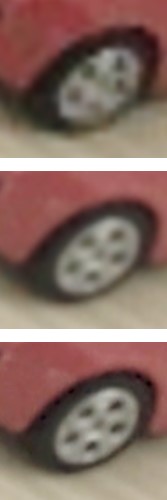}}~%
    \subfloat[]{\includegraphics[width=0.383\linewidth]{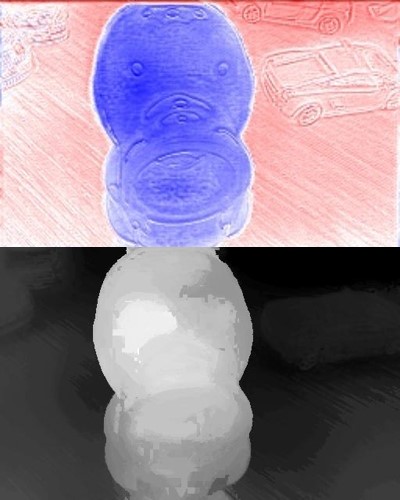}}~%
	\subfloat[]{\includegraphics[width=0.383\linewidth]{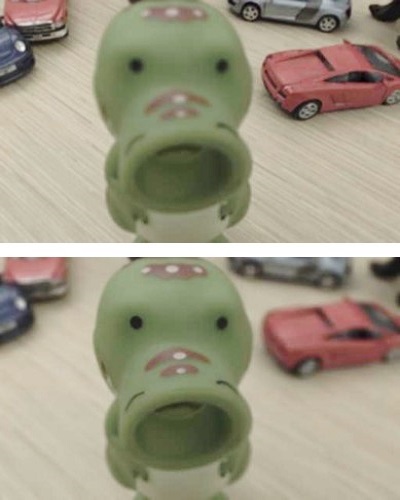}}~%
\end{center}
\caption{Preview of the proposed method. (a) Patches of spatial SR results (Input, 2$\times$, and 4$\times$). (b) Estimated appearance flow and depth image. (c) Refocused image at foreground and background.}
    \label{fig:Intro_Fig}
\end{figure}
At present, a light field image is captured using either plenoptic (light field) cameras~\cite{Raytrix_2013} or camera arrays~\cite{Wilburn_TOG05}.
However, the absence of the only available consumer light field camera, {\em i.e.} Lytro, has created a gap between consumers and light field experiences.
In addition, light field camera also suffers from spatial and angular resolution trade-off due to the sensor limitation.
Most light field camera favor angular resolution which results in a low spatial resolution light field image.
We focus on filling this gap so that end users can experience the advantages of light field imaging and beyond.
The idea is to jointly synthesize light field through angular and spatial super-resolution (SR) from only a single image which is abundantly available in the real world.
In order to do that, the geometry information from a single image should be inferred and used to synthesize the surrounding angular images.
Synthesizing a 4D light field is a severely ill-posed problem, but the impact of such work is considerably significant.
For example, promoting a single image into a densely sampled light field can elevate existing AR/VR immersion experiences.

In this context, light field synthesis has attracted considerable attention in recent years~\cite{Kalantari_TOG16,Mildenhall_TOG19,Srinivasan_ICCV17,Srinivasan_CVPR19,Wang_ECCV18,Wu_CVPR17,Yeung_ECCV18}.
Previous approaches can be grouped into two categories based on the input type, {\em i.e.} single or multi-view inputs.
The multi-view input utilizes multiple images captured from specific viewpoints to infer the geometric clue and use it to synthesize the light field.
However, only a few consumer cameras can simultaneously capture multi-view images, which makes the approach impractical for general use.

Existing method involving a single input utilizes two-stage neural networks and depth image-based rendering (DIBR) technique to synthesize the light field~\cite{Srinivasan_ICCV17}.
\cite{Srinivasan_ICCV17} is inspired by previous view synthesis techniques using geometry estimation~\cite{Flynn_CVPR16,Garg_ECCV16,Godard_CVPR17,Xie_ECCV16}.
However, \cite{Srinivasan_ICCV17} is highly dependent on the estimated depth quality and physical-based depth warping to synthesize angular images.
The depth-based approach also faces difficulty in reconstructing the occlusion and homogeneous region.
Typical learning based works rely on minimizing the error between the synthesized view and the ground truth image straightforwardly.
This leads the network to rely on pixel intensity cue and cannot be easily generalized to data with different and complex distribution.

In this paper, we develop a novel joint deep neural network for spatial and angular light field SR that utilizes the appearance flow to synthesize novel views.
To the best of our knowledge, this is the only work that tackle the joint super-resolution problem using only a single image.
We also introduce a spatio-angular consistent loss function based on known light field domain knowledge that helps the network learn robustly and efficiently.
Figure~\ref{fig:Intro_Fig} shows the result and application of the proposed method.

The key contributions of this paper are summarized as follows.
\begin{itemize}
\itemsep0.01em
\item End-to-end encoder-decoder style for joint spatial and angular light field SR model.
\item Novel spatio-angular consistent (light field based) loss that imposes geometric reasoning to the network.
\item Capability to be generalized to arbitrary scenes rather than a specific class of object compared with the previous approach.
\end{itemize}
\begin{figure*}[t]
\begin{center}
    {\includegraphics[width=1.0\linewidth]{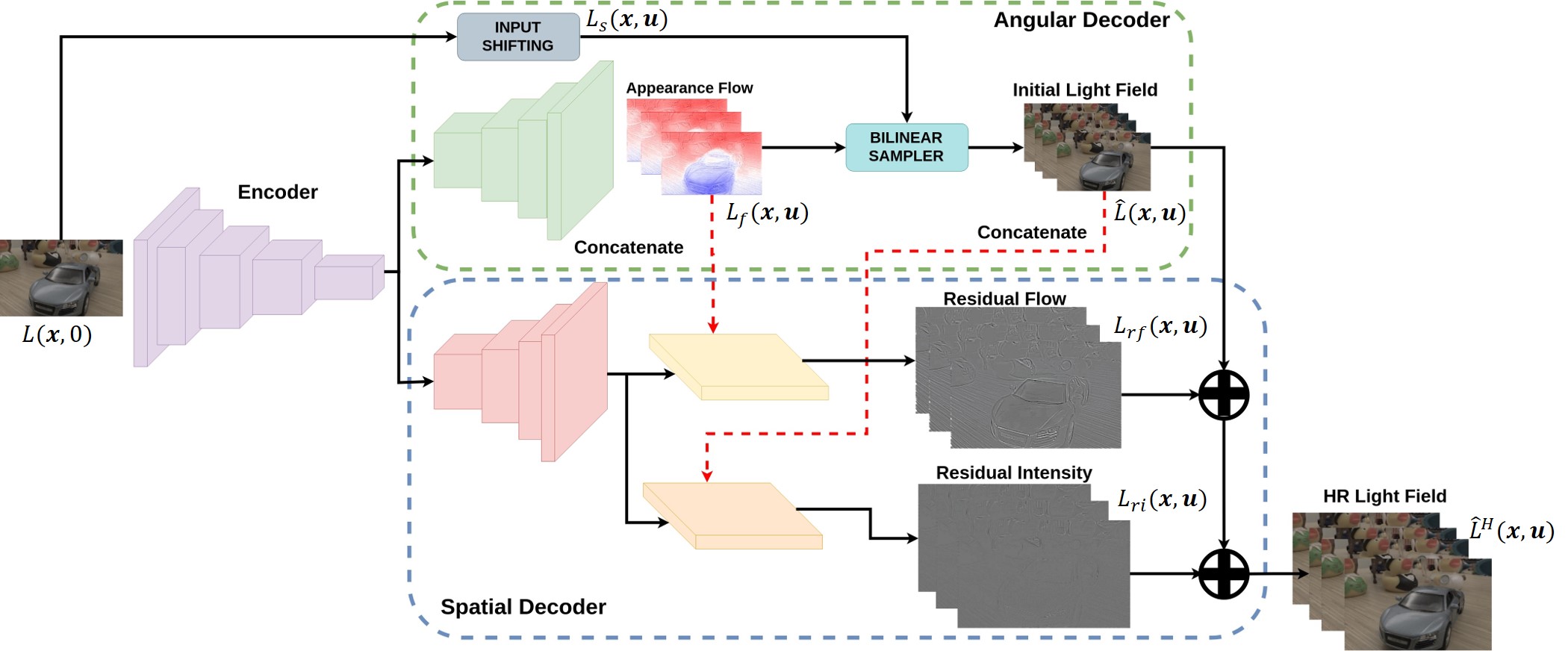}}
\end{center}
\caption{The structure of the proposed joint framework. Dashed green and blue lines denote the angular decoder and spatial decoder, respectively. Red dashed line denotes channel wise concatenation operation.}
    \label{fig:Network}
\end{figure*}
\section{Related Works}
Learning based light field synthesis (angular SR) has been investigated by many researchers in the past few years~\cite{Kalantari_TOG16,Mildenhall_TOG19,Srinivasan_CVPR19,Srinivasan_ICCV17,Wang_ECCV18,Wu_CVPR17,Yeung_ECCV18}.
On the basis of the number of input images, related approaches are categorized into multi-view image~\cite{Kalantari_TOG16,Wang_ECCV18,Wu_CVPR17,Yeung_ECCV18} and single image~\cite{Srinivasan_ICCV17} based.
Multi-view image-based light field synthesis is impractical due to its specific input pattern.
Meanwhile a single image-based light field synthesis is severely ill-posed although it is the most practical approach for light field synthesis.

\paragraph{\textbf{Sparse-Input Light Field Synthesis}}
Wanner and Goldluecke~\cite{Wanner_PAMI14} introduced a light field SR framework adopting the estimated depth information and variational optimization to fill missing views from a sparse light field image.
Phase-based light field synthesis from a micro-baseline stereo pair was proposed by Zhang~{\em et. al}~\cite{Zhang_CVPR15}.
Those studies were rooted on traditional approaches that use complex processing and various optimization approaches.
Meanwhile, learning-based view synthesis achieves better results by using an end-to-end training strategy.
Zhou~{\em et. al}~\cite{Zhou_ECCV16} proposed a new geometric representation called appearance flow to synthesize an image with a novel view.
However, the proposed representation is not generalized well to a complex scene with multiple object and non-homogeneous background.
Zhou~{\em et. al}~\cite{Zhou_TOG18} presented a novel geometric representation called multi-plane images~(MPI) to synthesize a horizontal light field from a narrow baseline camera.
Srinivasan~{\em et. al}~\cite{Srinivasan_CVPR19} extended the MPI extrapolation boundaries based on the Fourier domain analysis.
Recent work by Mildenhall~{\em et. al}~\cite{Mildenhall_TOG19} exhibited state-of-the-art view synthesis performance with multi MPI and a blending technique.
However, these approaches require camera pose information and/or multiple inputs to synthesize the novel view which is not commonly available in the real world.
In addition, MPI requires a significant computing resource(increasing the layer count) to get an accurate result.

Kalantari~{\em et. al}~\cite{Kalantari_TOG16} introduced the first learning based light field synthesis solution.
They utilized four corner images to synthesize a 4D light field using a depth estimation, warping, and color refinement approach.
The inputs to the depth estimation network were mean and variance images, as inspired by the depth estimation work of~\cite{Tao_ICCV13}.
Wu~{\em et. al}~\cite{Wu_CVPR17} utilized an epipolar plane image~(EPI) obtained from sparse input images and synthesized an up-sampled EPI through a specially designed blur kernel.
The framework is then extended further into several applications~\cite{Wu_PAMI19}.
Wang~{\em et. al}~\cite{Wang_ECCV18} employed a pseudo 4DCNN represented as 2D strided convolution and 3DCNN, where the light field image was synthesized in a step-by-step manner.
Yeung~{\em et. al}~\cite{Yeung_ECCV18} applied a high dimensional convolutional kernel to infer spatial and angular information from sparse input images.
In summary, sparse input light field synthesis focuses on synthesizing in-between views and could be regarded as solving an interpolation problem.
The specific input sampling pattern also hinders its practical usage.

\paragraph{\textbf{Single-Input Light Field Synthesis}}
Srinivasan~{\em et. al}~\cite{Srinivasan_ICCV17} introduced the first solution to solve light field synthesis from a single image.
They proposed a single image based depth estimation to obtain the approximate geometry of a scene.
Then, the estimated depth was utilized to synthesize a novel view using the DIBR approach.
However, their method is constrained to a simple scene and highly dependent on pixel intensity.

In this paper, we focus on solving the problem of single-image light field angular synthesis.
Contrary to~\cite{Srinivasan_ICCV17}, we propose to use an alternative geometric representation (appearance flow) and present a light field based loss function to avoid reliance on bright pixel color.
Furthermore, we go beyond the problem scope of~\cite{Srinivasan_ICCV17} and solve the spatial resolution problem simultaneously.

\section{Proposed Method}
\subsection{Joint Light Field Super-Resolution }
This paper aims to synthesize a high resolution 4D light field ${L}(\textbf{x},\textbf{u})$ given a single image that serves as the central sub-aperture image~(SAI) ${L}(\textbf{x},\textbf{0})$.
We follow the two-plane parametrization of light field ${L}(\textbf{x},\textbf{u})$, introduced by~\cite{Levoy_CGI96}, where $\textbf{x}$ and $\textbf{u}$ are the coordinates in spatial and angular planes, respectively.
In general, the light field synthesis problem is formulated as DIBR problem, which is described as follows:
\begin{equation}
{L}(\textbf{x},\textbf{u}) = {L}(\textbf{x} + d(u), 0),
\end{equation}
where $d(u)$ is the disparity in $x$ direction.
Disparity depends on the depth information of the central image and the novel angular coordinate $u$.
We address the light field synthesis problem by using an approximation function $f(\cdot)$ represented as a deep convolutional neural network, as described in
\begin{equation}
{L}(\textbf{x},\textbf{u}) = f({L}(\textbf{x},\textbf{0})).
\end{equation}
Function $f$ solves a highly ill-posed problem.
To solve the problem jointly, we design an encoder and multi decoder framework.
We use the shared encoded feature to solve both angular and spatial light field SR.
The joint framework is decomposed into two decoder branches.
The top (angular) branch solves the angular SR problem.
While the bottom (spatial) branch solves the spatial SR problem using additional information estimated by the first decoder branch.
The overall network structure is shown in Figure~\ref{fig:Network}.

\paragraph{\textbf{Angular Decoder}}
The angular decoder branch estimates appearance flows to extrapolate the central view to each SAI in the 4D light field.
Appearance flow~\cite{Zhou_ECCV16} represents 2D coordinate vectors specifying where pixels are mapped in the reconstructed novel views.
Appearance flow is accompanied with little blur, preserves color identities, and removes the dependency on the physical-based approach to synthesize novel views.
Considering that the ground truth light field appearance flow is difficult and expensive to obtain, the proposed network is designed to estimate appearance flow in an unsupervised manner.
The network learns to estimate appearance flows by supervising the synthesized light field (warped novel views) image.

The angular decoder is decomposed into three sub-problems, {\em i.e.} appearance flow estimation for each viewpoint $u$, image shifting with respect to the central view, and novel view extrapolation.
Each sub-problem can be defined mathematically as follows.
\begin{align}
{L}_f(\textbf{x},\textbf{u}) &= \mathcal{F}({L}(\textbf{x},\textbf{0})) \\
{L}_s(\textbf{x},\textbf{u}) &= \mathcal{S}({L}(\textbf{x},\textbf{0}), \nabla(u)) \\
\hat{L}(\textbf{x},\textbf{u}) &= \mathcal{W}({L}_s(\textbf{x},\textbf{u}), {L}_f(\textbf{x},\textbf{u})),
\end{align}
where $\mathcal{F}$ estimates appearance flow ${L}_f$ for each novel angular view.
$\mathcal{S}$ performs angular shifting to the position $\nabla(u)$ of novel views.
Image shifting serves as a bias initialization for the network.
$\mathcal{W}$ is the warping function for shifted image ${L}_s(\textbf{x},\textbf{u})$ using its corresponding appearance flow ${L}_f(\textbf{x},\textbf{u})$.
Image warping is performed using a bilinear sampler module~\cite{Jaderberg_NIPS15} to produce the light field image $\hat{L}(\textbf{x},\textbf{u})$.

\paragraph{\textbf{Spatial Decoder}}
We solve the angular SR first followed by spatial SR jointly.
The spatial decoder branch estimates a residual light field image to enhance the initially estimated 4D light field (angular SR).
The final HR light field is the result of adding multi-stage residual images.
The purpose of multi-stage residual estimation is to solve missing high-frequency information and to post-process the initial light field.
They are further discussed in Section~\ref{sec:SpatialSR}.
The problem can be defined as follows.
\begin{equation}
{L}_{rf}(\textbf{x},\textbf{u}), {L}_{ri}(\textbf{x},\textbf{u}) = \mathcal{R}({L}_f(\textbf{x},\textbf{u}), \hat{L}(\textbf{x},\textbf{u}))
\end{equation}
\begin{equation}
\hat{L}^H(\textbf{x},\textbf{u}) = \hat{L}(\textbf{x},\textbf{u}) + {L}_{rf}(\textbf{x},\textbf{u}) + {L}_{ri}(\textbf{x},\textbf{u})
\end{equation}
where $\mathcal{R}$ outputs two residual light field images, namely residual flow ${L}_{rf}(\textbf{x},\textbf{u})$ and residual intensity ${L}_{ri}(\textbf{x},\textbf{u})$.
The output of spatial decoder layer is concatenated with appearance flow and initial light field image, indicated by red dashed lines in Figure~\ref{fig:Network}.
Residual flow and residual intensity are named based on the prior information concatenated to produce the residual light field.
The final high resolution light field is denoted by $\hat{L}^H(\textbf{x},\textbf{u})$.

The proposed objective function is defined as
\begin{equation}
\begin{aligned}
\underset{\theta}{\text{min}}
    \sum[
            \lambda{_g}L_{g}(\theta)+
            \lambda{_{l}}L_{l}(\theta) +
            \lambda{_{tv}}\psi_{tv}(\theta) +
            \lambda{_{sr}}L_{sr}(\theta)
            ],
\end{aligned}
\end{equation}
where $\theta$ denotes the deep neural network parameters.
The problem formulation and objective function enable the network to estimate an appearance flow for each SAI at $u$ in an unsupervised manner through the supervision of synthesized pixels.
The common pixel-wise loss is not utilized in the proposed method because it does not enforce the geometry constraint to the network.
Instead, we rely on the known light field domain knowledge and design two geometrically constrained losses, {\em i.e.} global $L_{g}(\theta)$ and local $L_{l}(\theta)$ light field losses.
Both loss functions are useful in preserving the spatio-angular consistency between light field SAIs.
In addition, we propose a regularization loss $\psi_{tv}(\theta)$ to the estimated appearance flow to enforce angular consistency.
$L_{sr}(\theta)$ denotes the spatial SR loss.
Each loss is weighted with a corresponding $\lambda$ to balance each loss contribution accordingly.

\begin{figure}[t]
\centering
    \subfloat[]{\includegraphics[width=0.33\linewidth]{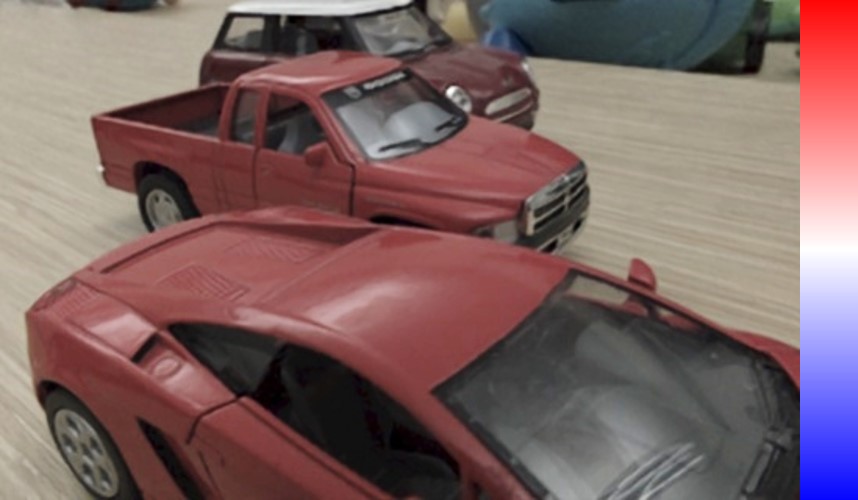}}~%
    \subfloat[]{\includegraphics[width=0.30\linewidth]{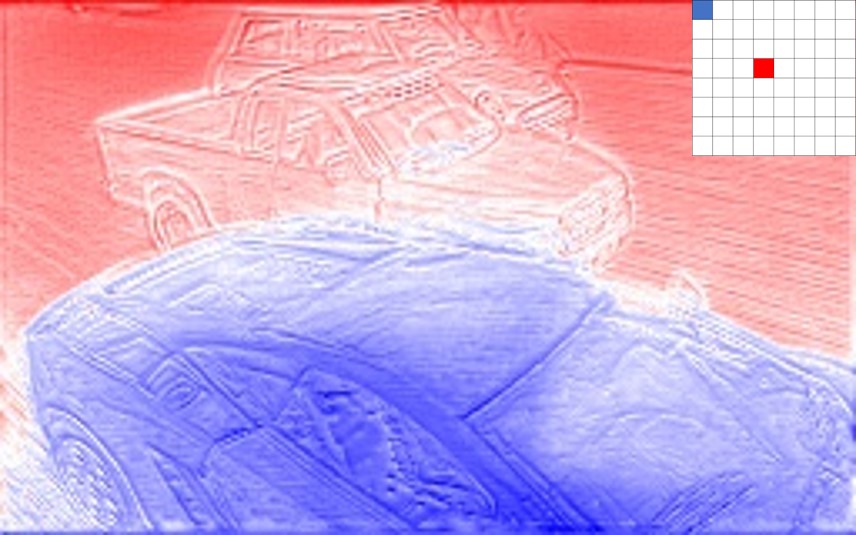}}~%
    \subfloat[]{\includegraphics[width=0.30\linewidth]{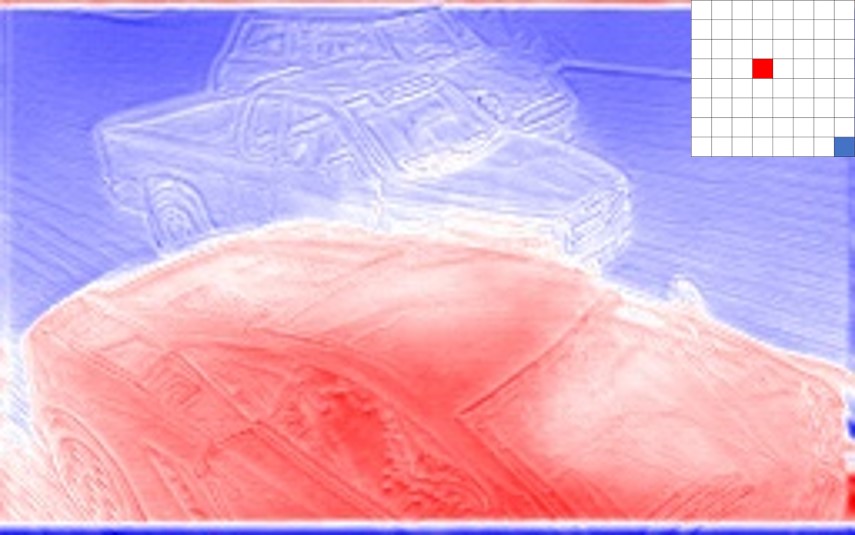}}~%
\caption{Estimated appearance flow. (a) Input image. (b) Appearance flow for top left corner. (c) Appearance flow for bottom right corner. The color show the direction of the flow with white color denotes zero flow.}
    \label{fig:Appearance_Flow}
\end{figure}
\subsection{Angular Super-Resolution}
Angular SR decoder relies on estimating appearance flow to warp the shifted image into novel SAIs.
To estimate an accurate and dense appearance flow, we designed the encoder to extract all important representation in the image.
Sharing encoded feature with both decoders encourages the encoder to extract important features in the input image.
Angular decoder consists of convolution layers with skip connections followed by leaky ReLU~\cite{Maas_ICML13}.
Additional details about the network structure are available in the supplementary material.

In particular, the proposed flow estimation network produces coordinate vectors $(x,y)$ to sample from ${L}_s(\textbf{x},\textbf{u})$ using the bilinear sampler to synthesize the novel view $\hat{L}(\textbf{x},\textbf{u})$.
The appearance flow is visualized in Figure~\ref{fig:Appearance_Flow}.
It is observed that the estimated appearance flow is smooth, edge-aware, and consistent in the inverse direction.
In addition, it robustly estimates the geometry of a scene with similar intensities.
We learn $\mathcal{F}$ to estimate the appearance flow by minimizing the loss between the extrapolated novel views $\hat{L}(\textbf{x},\textbf{u})$ and ground truth light field ${L}(\textbf{\textbf{x}},\textbf{u})$.
The angular decoder estimates appearance flow for all novel SAIs in a single shot.

Image shifting is designed to guide the network by providing an initial bias.
This technique is inspired by the work of Xie~{\em et. al}~\cite{Xie_ECCV16}, in which the input image is shifted to guide the network in synthesizing the corresponding stereo image.
The shifting operation can be written as follows.
\begin{equation}
{L}_s(\textbf{x},\textbf{u}) = {L}(\textbf{x}-\eta \Delta u, \textbf{0}),
\end{equation}
where $\eta$ is the constant angular shift value in horizontal and vertical directions.
$\Delta u$ is the angular distances between novel and central views.
Considering the redundancy in light field SAI, we can partially imitate how pixels shift to each angular position and utilize this to provide better initialization.
The $\eta$ value is predetermined based on the disparity between SAI in the target light field.
\begin{figure*}[t]
\centering
	{\includegraphics[width=0.33\linewidth]{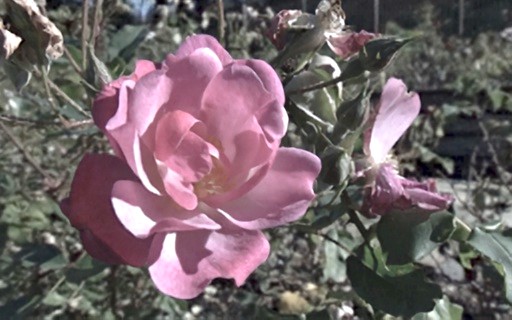}}~%
    {\includegraphics[width=0.33\linewidth]{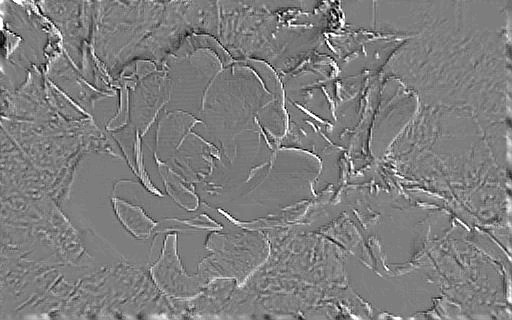}}~%
    {\includegraphics[width=0.33\linewidth]{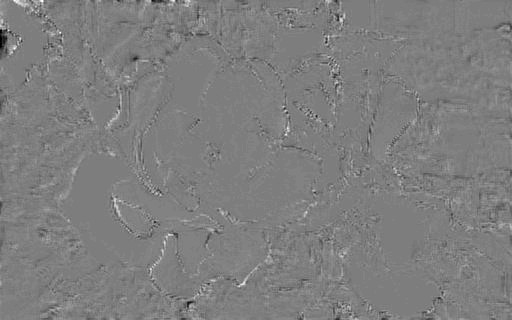}}

    \vspace{-1.0mm}
    \subfloat[]{\includegraphics[width=0.33\linewidth]{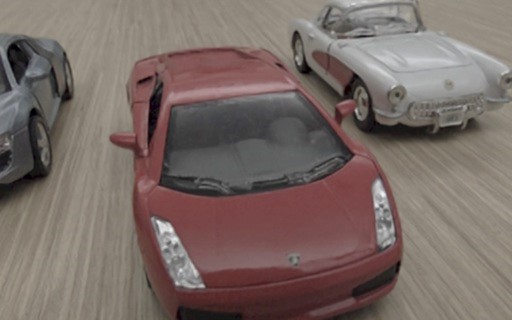}}~%
    \subfloat[]{\includegraphics[width=0.33\linewidth]{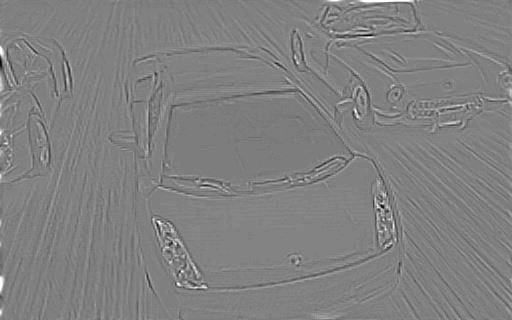}}~%
    \subfloat[]{\includegraphics[width=0.33\linewidth]{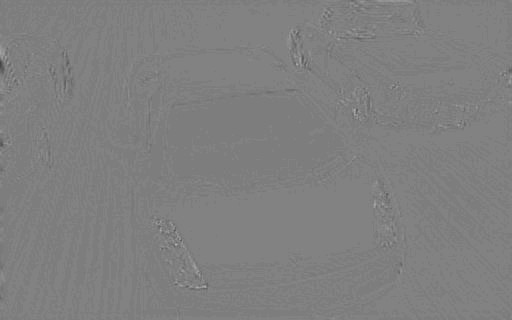}}~%
\caption{Residuals estimated from concatenating appearance flow and initial light field image. (a) Input image. (b) Residual flow, ${L}_{rf}(x,u)$. (c) Residual intensity, ${L}_{ri}(x,u)$.}
    \label{fig:Residual_Visualization}
\end{figure*}
\subsection{Spatial Super-Resolution}
\label{sec:SpatialSR}
In spatial SR, we estimate residual images to recover the high frequency information which is lost during initial light field upsampling.
Moreover, we try to refine the initial light field image.
The spatial decoder is designed to solve this problem by adding the initial light field image with residual images.
In particular, we incorporate information estimated by angular decoder into the spatial decoder, such as appearance flow and the initial light field image.
We concatenate those information before the final convolution layer in the decoder.

The estimated residual image confirms the effectiveness of the proposed multi-stage residual framework, as shown in Figure~\ref{fig:Residual_Visualization}.
The first stage estimate the high frequency information, such as edge region and textured (details) region.
While the second stage focus on a more sparse estimation of occlusion and erroneous region.
The second stage can be seen as post-processing or refinement part of the framework.

The spatial decoder estimates residual image for every SAIs in the light field image in a single shot.
During inference, given a single input, the network estimates high resolution light field in a single run.
While in the training stage, the spatial decoder is frozen for several iterations before both decoders are trained together in an end-to-end fashion.
These are discussed further in Section~\ref{sec:Experiment}.
\subsection{Light Field Loss}
\label{sec:LFLoss}
Although the proposed framework enables us to solve the super-resolution problem jointly, a good objective function for learning the relation between angular views is a mandatory.
L1 or L2 loss, which are commonly used by conventional approaches, cannot provide proper geometric reasoning to the network.
It encourages the network to look at dominant pixel color individually instead of understanding the whole scene.
We propose to use light field angular information instead of pixel information individually.
In specific, we use mean and variance of a light field image which exists in any light field image.

\paragraph{\textbf{Global Light Field Loss}}
We propose a novel 4D light field loss, which is formulated as
\begin{equation}
L_g(\theta) = |M(\hat{L}^\theta) - M({L})| + |V(\hat{L}^\theta) - V({L})|_1
\end{equation}
where
\begin{flalign}
M(L) = & \frac{1}{N}\displaystyle\sum_{s=1}^{N}L(\textbf{x},\textbf{u}_s) \\
V(L) = & {\frac{1}{N-1}\displaystyle\sum_{s=1}^{N}(L(\textbf{x},\textbf{u}_s)- M(L))^2}.
\end{flalign}
$M(L)$ and $V(L)$ denote the mean and variance while $s$ denotes the index of SAI in the light field.
Computing the light field mean is equivalent to obtaining the refocus image at zero disparity.
The refocused image correlates to the depth of the light field image.
We did not need labelled light field depth because it is embedded in the light field itself through light field mean.
Therefore, the synthesized light field depth can be explicitly evaluated in an efficient way and unsupervised manner.
Meanwhile, variance captures the difference between SAIs and helps the network learn the occlusion and edge region.
This is known from the light field depth estimation work of~\cite{Tao_ICCV13,Williem_CVPR16,Williem_PAMI17}, which utilize the mean and variance to compute defocus and correspondence responses, respectively.
Kalantari~{\em et. al}~\cite{Kalantari_TOG16} also employed the mean and variance images as the input to their depth estimation network.
In this paper, we show that mean and variance images can be used as loss function to help the network learn the light field geometry efficiently.
\begin{figure}[t]
\centering
    \includegraphics[width=1.0\linewidth]{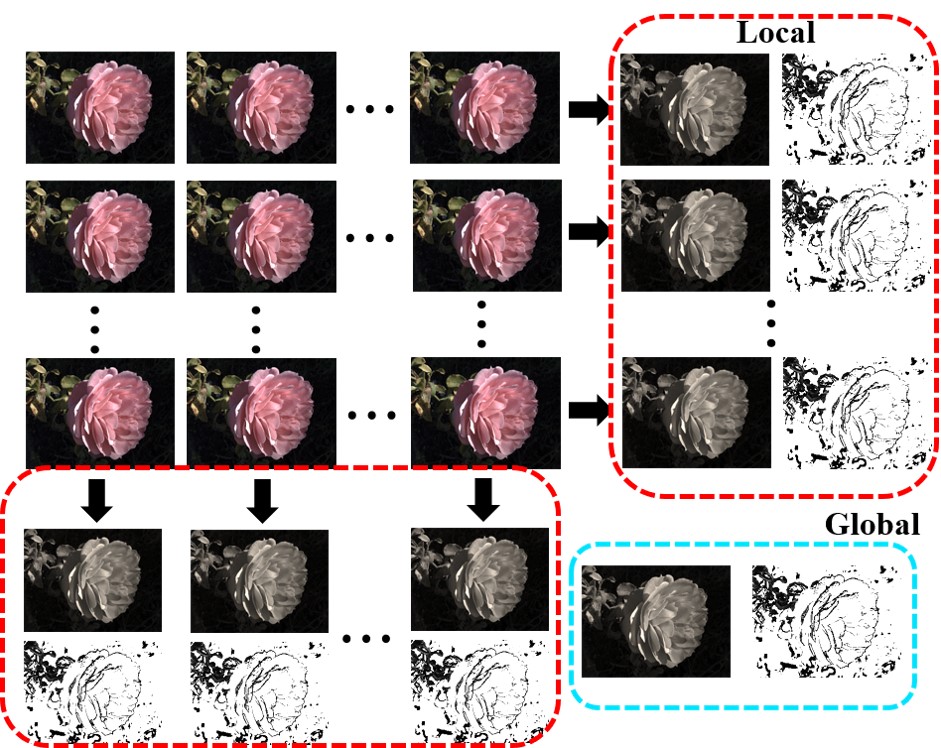}~%
\caption{Global and local light field losses. Blue and red dashed line denote the global and local losses, respectively.}
    \label{fig:LF_Loss}
\end{figure}
\begin{figure*}[t]
\centering
    \subfloat[]{\includegraphics[width=0.33\linewidth]{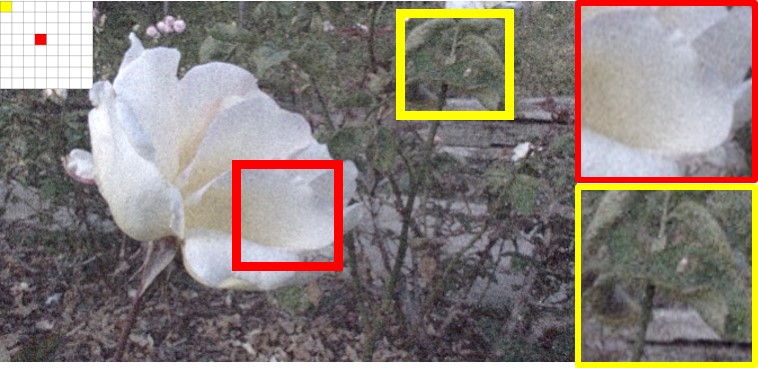}}~%
    \subfloat[]{\includegraphics[width=0.33\linewidth]{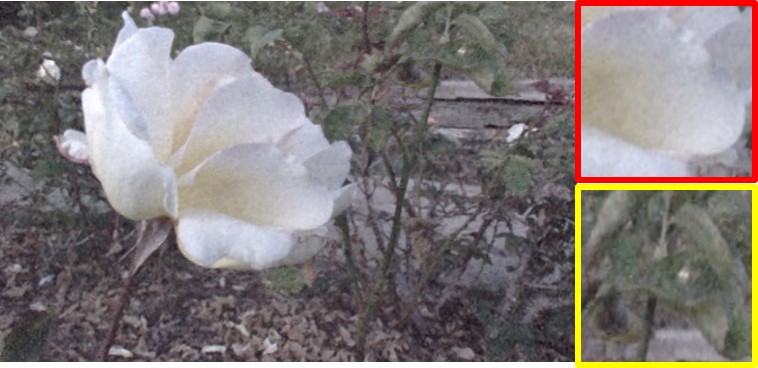}}~%
    \subfloat[]{\includegraphics[width=0.33\linewidth]{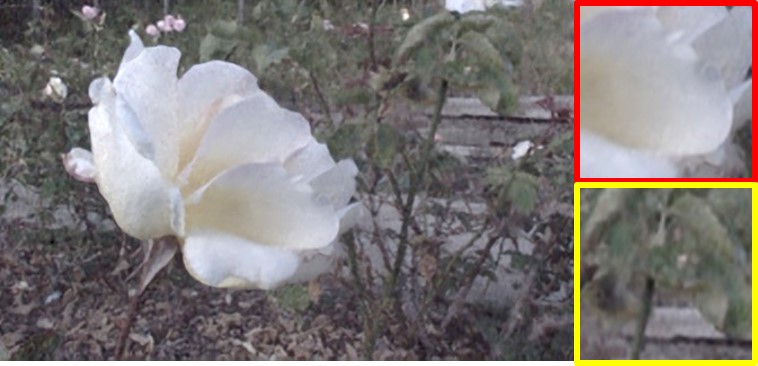}}~%

\caption{Denoising capability. (a) Noisy input. (b) Srinivasan~{\em et. al}~\cite{Srinivasan_ICCV17} result. (c) Proposed result. Srinivasan's result still contain some degree of noise.}
    \label{fig:Total_Variation}
\end{figure*}

\paragraph{\textbf{Local Light Field Loss}}
Although 4D global loss captures geometric information globally, the network should learn the local geometric relation between SAIs in a refined manner.
The idea is to help the network explicitly understand angular relation in the horizontal and vertical directions.
We compute the mean and variance for SAIs in each row and column in the 4D light field.
The losses at each light field row and column are accumulated to obtain the final local loss.
The process can be formulated as
\begin{equation}
\begin{aligned}
 L_e(\theta) = &\displaystyle\sum_{m=1}^{U} \displaystyle\sum_{n=1}^{U} |M(\hat{L}_{m,n}^\theta(\textbf{x},\textbf{u}_s)) - M({L_{m,n}}(\textbf{x},\textbf{u}_s))|_1 + \\
&|V(\hat{L}^\theta_{m,n}(\textbf{x},\textbf{u}_s)) - V({L_{m,n}}(\textbf{x},\textbf{u}_s))|_1,
\end{aligned}
\end{equation}
where $m,n$  denote the angular resolution of the light field image, and mean and variance are computed for $s \in \{1,\dotso,U\}$.
Without the loss of generality, we assume the light field angular resolution is equal in both horizontal and vertical directions.
Figure~\ref{fig:LF_Loss} visualizes the losses by computing the light field mean and variance.

\paragraph{\textbf{Loss Regularization}}
An inconsistent appearance flow might appear and cause artifacts between SAIs.
This problem is expected because appearance flow is estimated from a single image in an unsupervised manner.
An alternative approach is to use the conventional flow estimation method into the ground truth light field and compare it with the estimated flow.
However, this approach is tedious and increases framework complexity.
Thus, we present a strategy to remedy inconsistent and incorrect flow by incorporating a regularization term into the loss function.

To compress artifacts from inconsistent appearance flow, total variation regularization is applied.
Total variation is commonly used for noise removal in image processing.
The idea is to smooth inconsistent (noisy) appearance flow while keeping important edge information.
$L_2$ minimization is performed on the gradient of the estimated appearance flow.
\begin{equation}
\psi_{tv}(\theta)= ||\nabla_{x} \hat{L}^\theta(\textbf{x},\textbf{u})||_2.
\end{equation}
Moreover, total variation denoising capability is also carried on to the network, as shown in Figure~\ref{fig:Total_Variation}.

\paragraph{\textbf{Spatial Super-Resolution Loss}}
Spatial SR loss is straightforward.
We minimize the error of upsampled initial light field added by estimated residual images and the loss computation is defined as follows.
\begin{equation}
L_{sr}(\theta) = |\hat{L}^H(\textbf{x},\textbf{u}) - {L}^H(\textbf{x},\textbf{u})|_1 ,
\end{equation}
where ${L}^H(\textbf{x},\textbf{u})$ is the ground truth high resolution light field image.
We upsample the initial light field image using bilinear interpolation.
\begin{figure*}
\begin{center}

    {\includegraphics[width=0.31\linewidth]{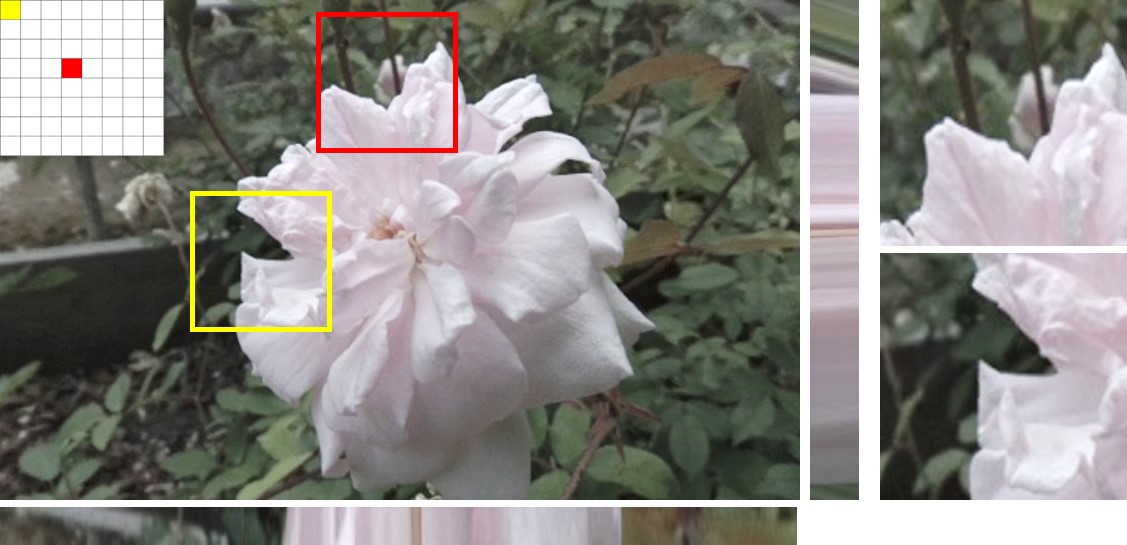}}~%
    {\includegraphics[width=0.31\linewidth]{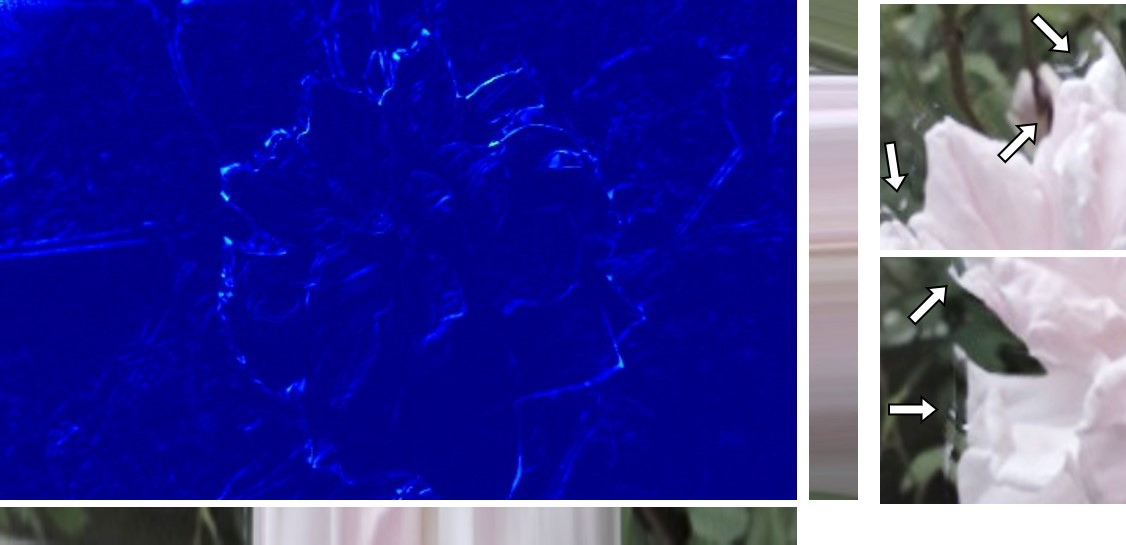}}~%
    {\includegraphics[width=0.31\linewidth]{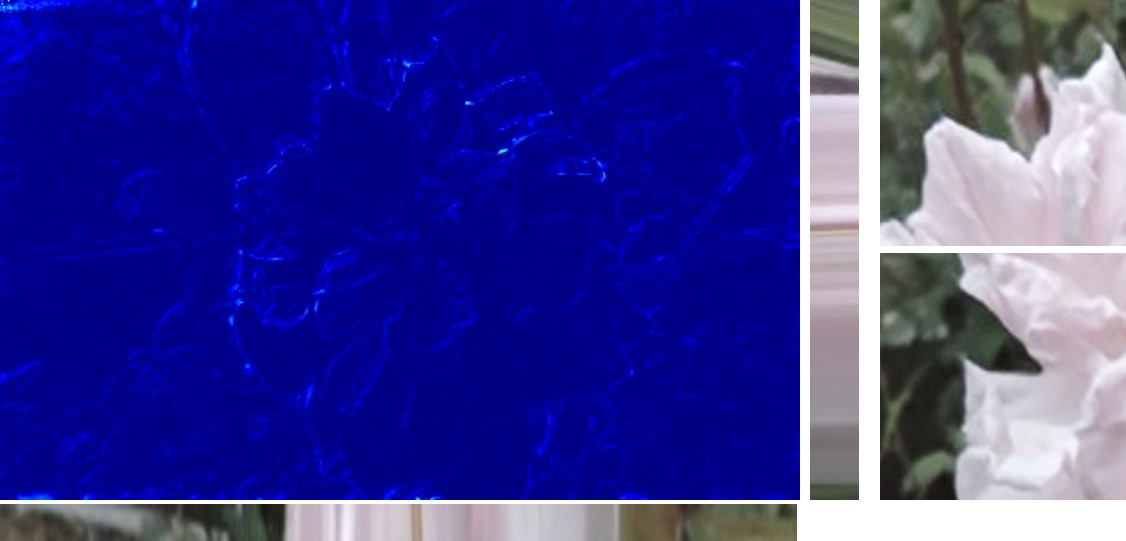}}~%

    \vspace{1.5mm}
    {\includegraphics[width=0.31\linewidth]{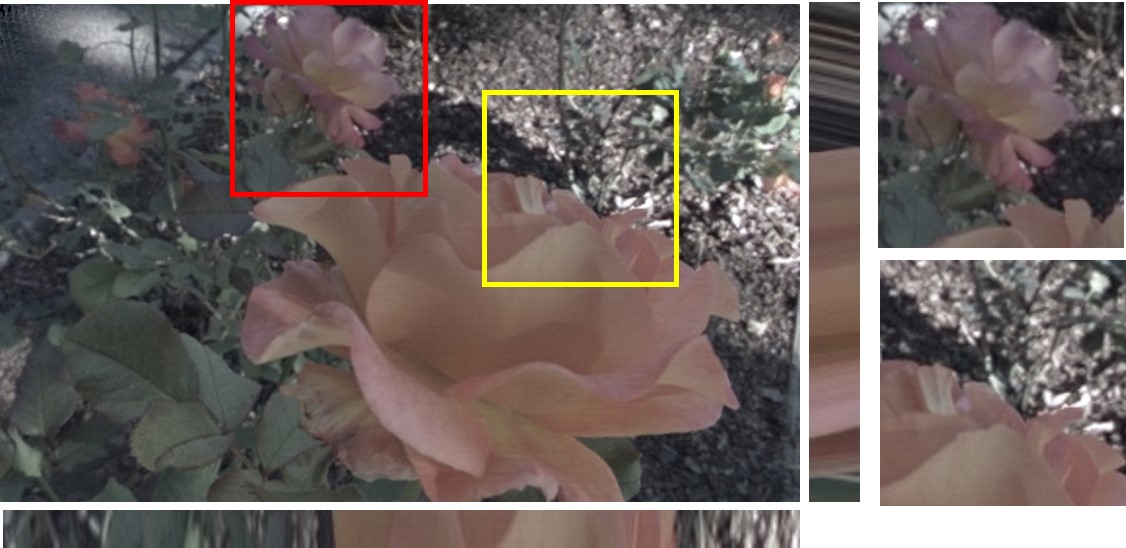}}~%
    {\includegraphics[width=0.31\linewidth]{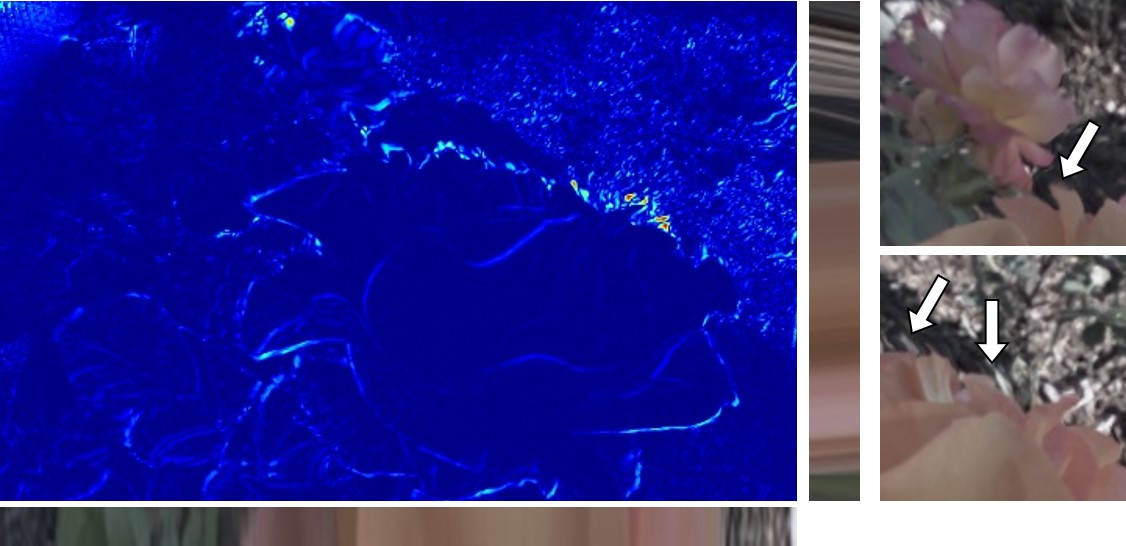}}~
    {\includegraphics[width=0.31\linewidth]{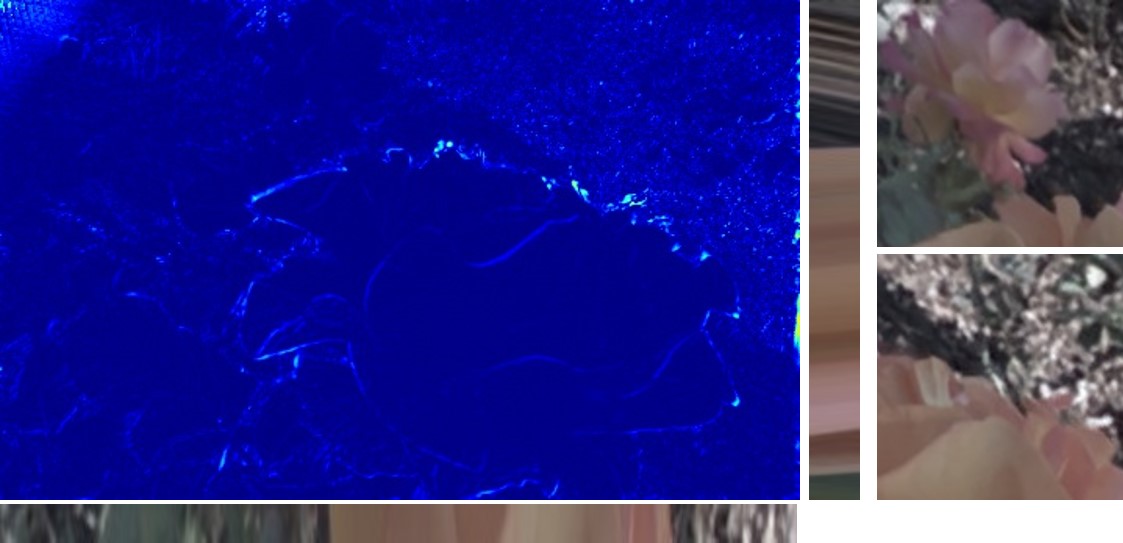}}~%

    \vspace{1.5mm}
    {\includegraphics[width=0.31\linewidth]{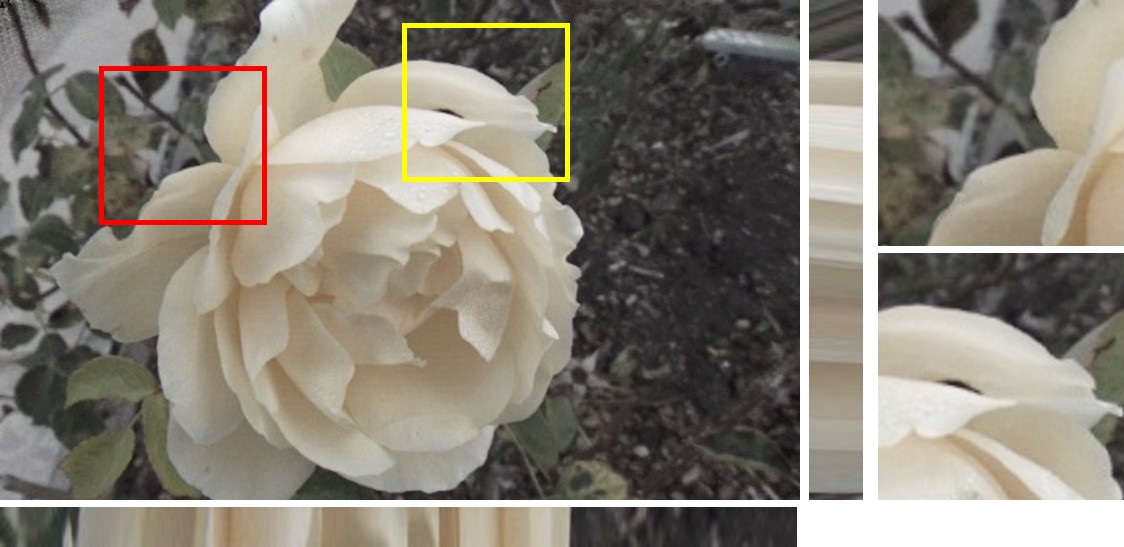}}~%
    {\includegraphics[width=0.31\linewidth]{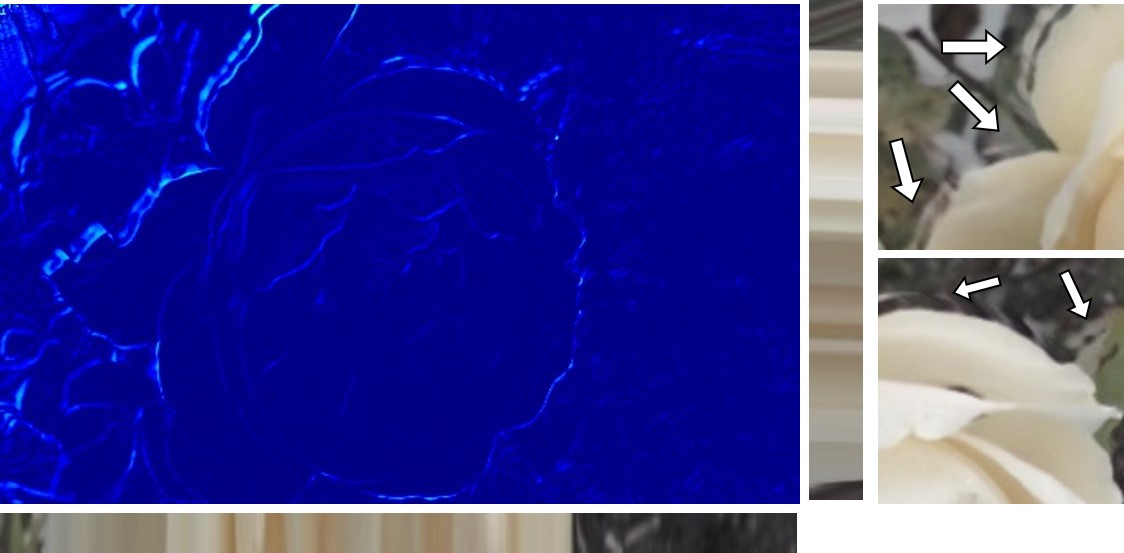}}~%
    {\includegraphics[width=0.31\linewidth]{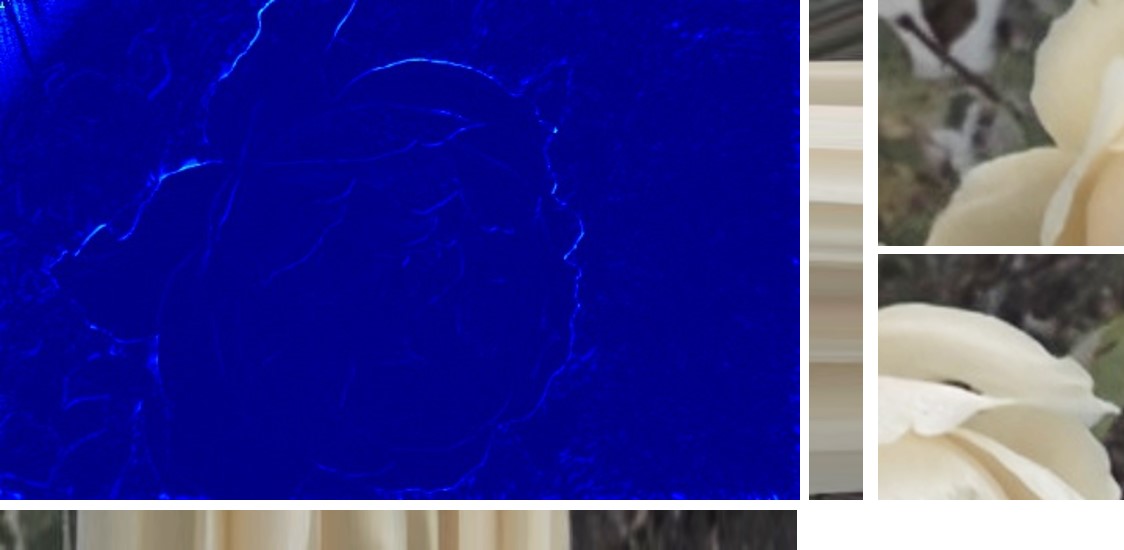}}~%

    \vspace{1.5mm}
    {\includegraphics[width=0.31\linewidth]{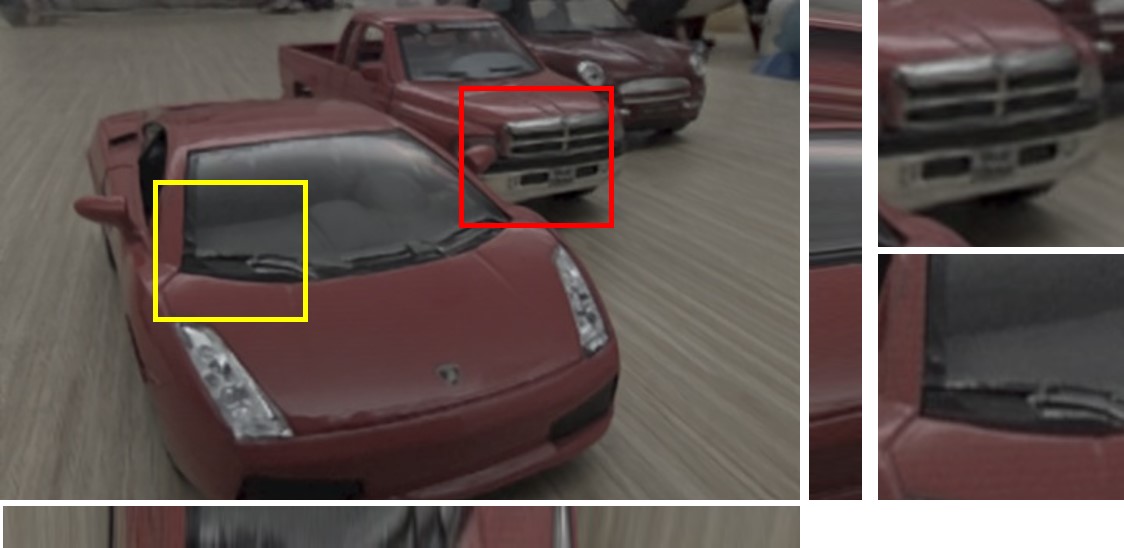}}~%
    {\includegraphics[width=0.31\linewidth]{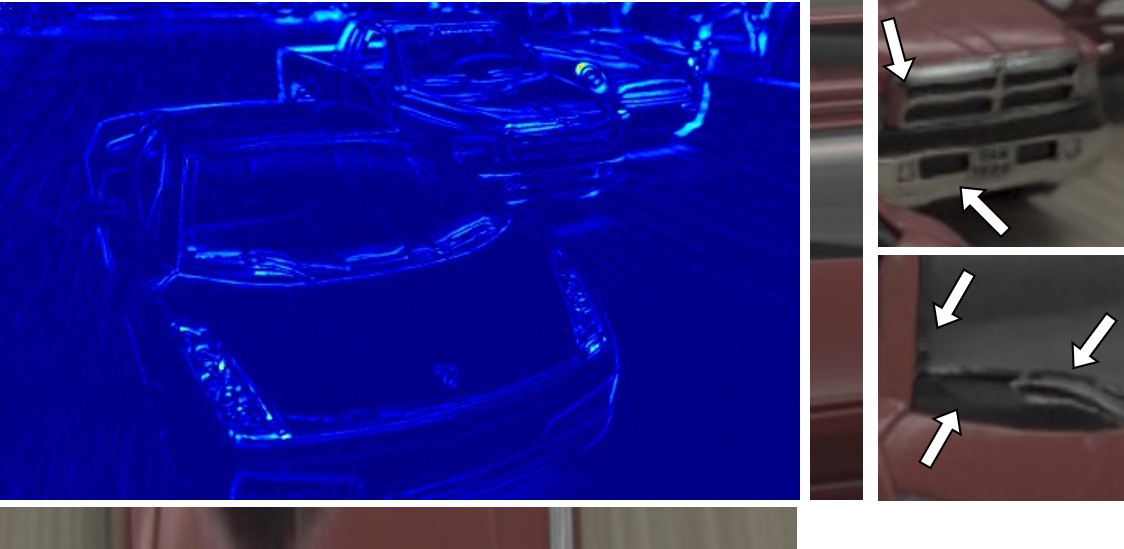}}~%
    {\includegraphics[width=0.31\linewidth]{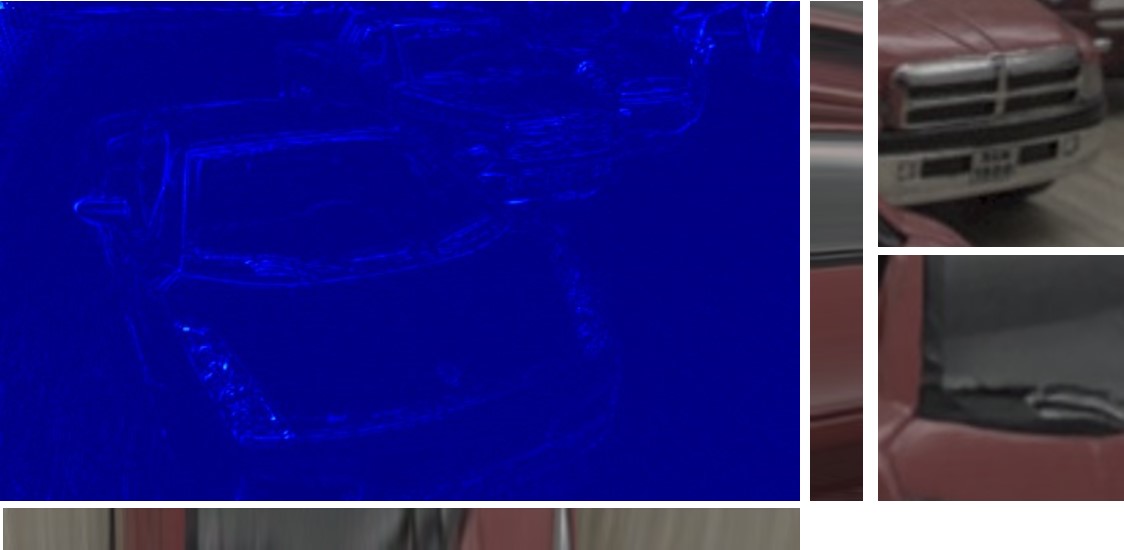}}~%

    \vspace{1.5mm}
    {\includegraphics[width=0.31\linewidth]{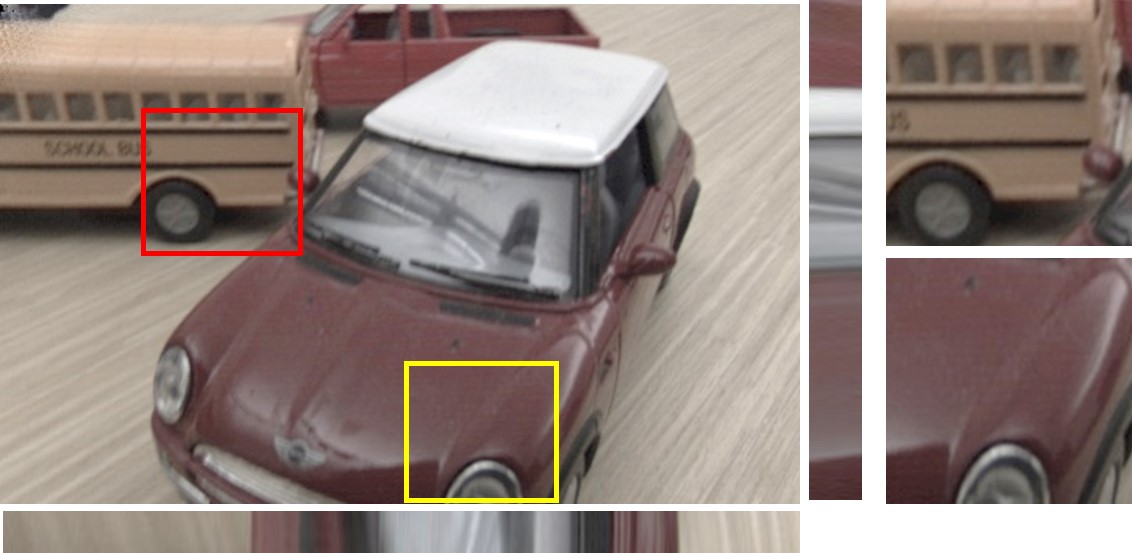}}~%
    {\includegraphics[width=0.31\linewidth]{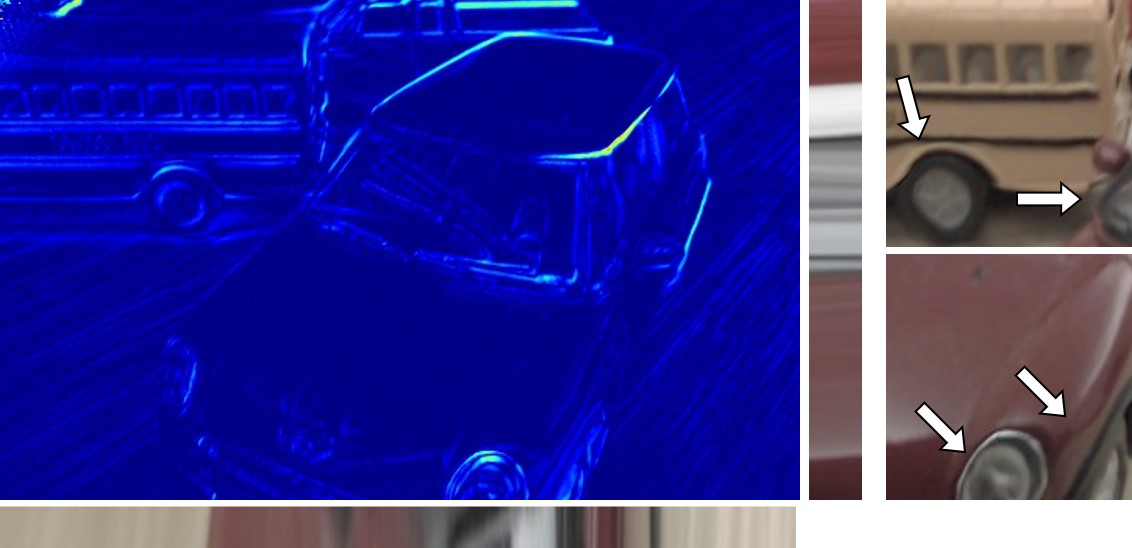}}~%
    {\includegraphics[width=0.31\linewidth]{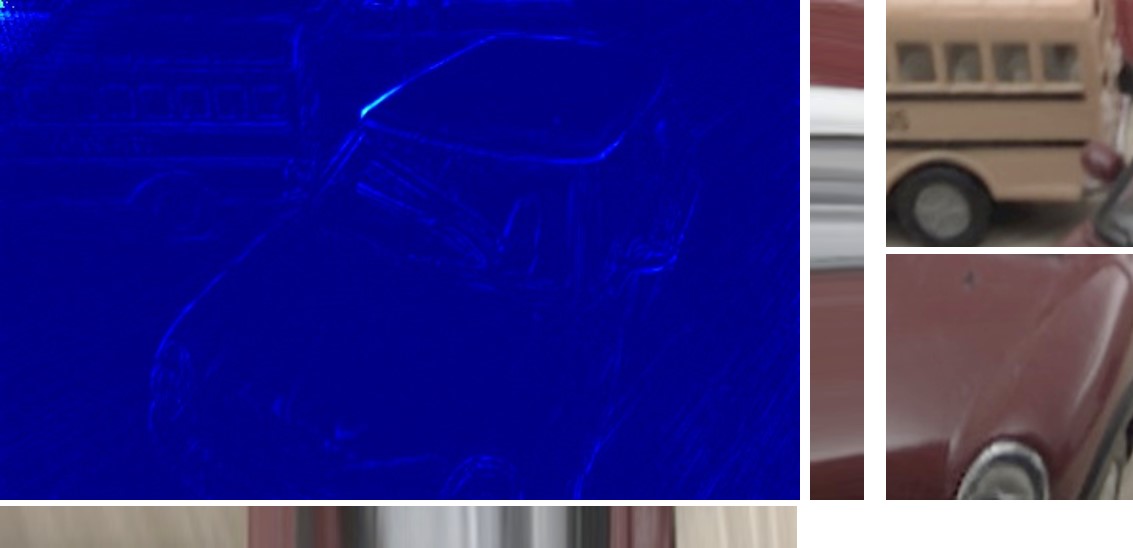}}~%

    \vspace{1.0mm}
	\stackunder[5pt]{\includegraphics[width=0.31\linewidth]{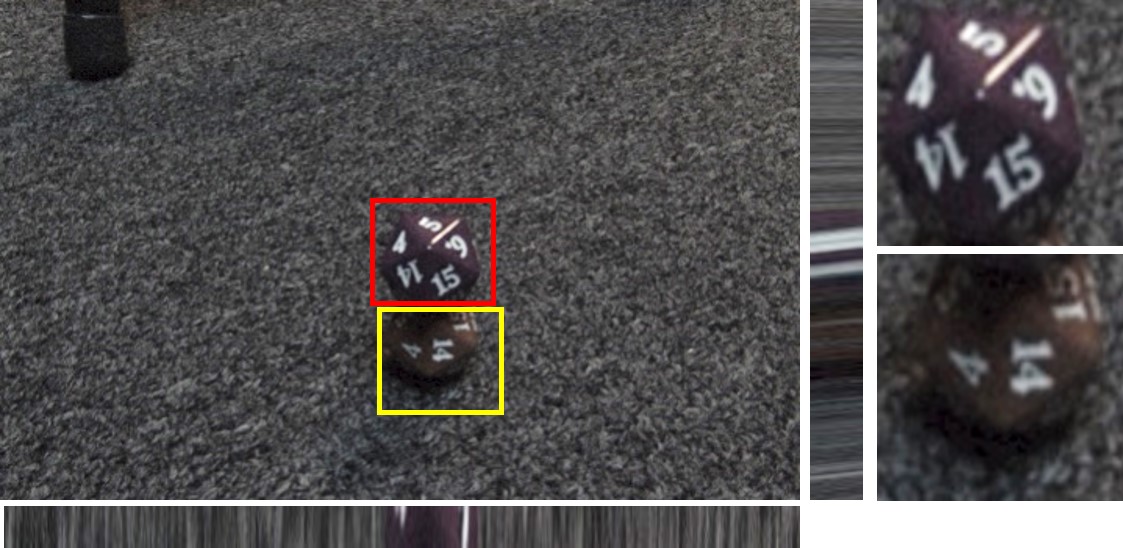}}{Reference image}~%
    \stackunder[5pt]{\includegraphics[width=0.31\linewidth]{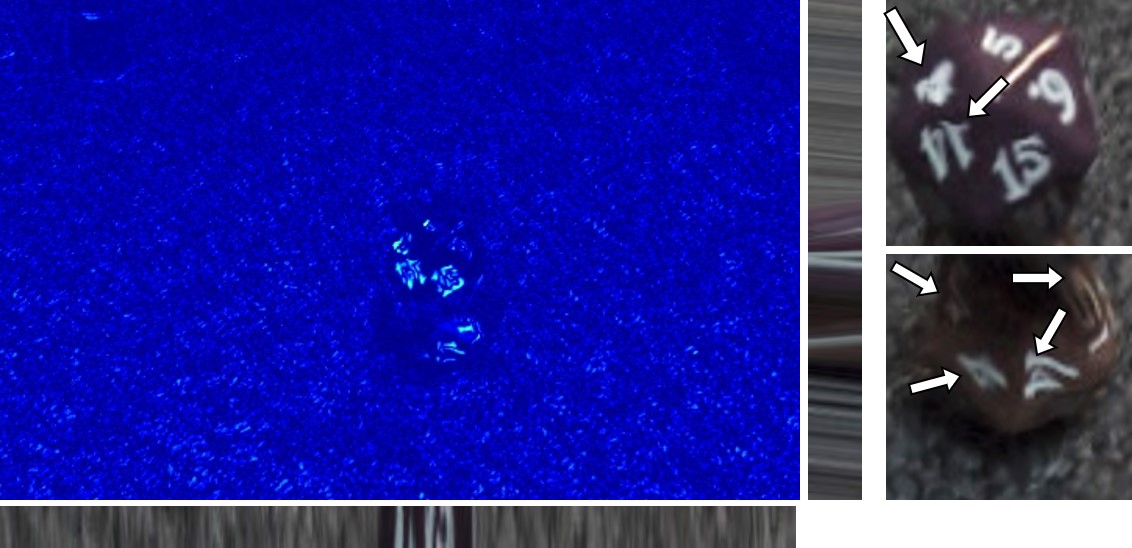}}{Srininvasan~{\em et. al}~\cite{Srinivasan_ICCV17}}~%
	\stackunder[5pt]{\includegraphics[width=0.31\linewidth]{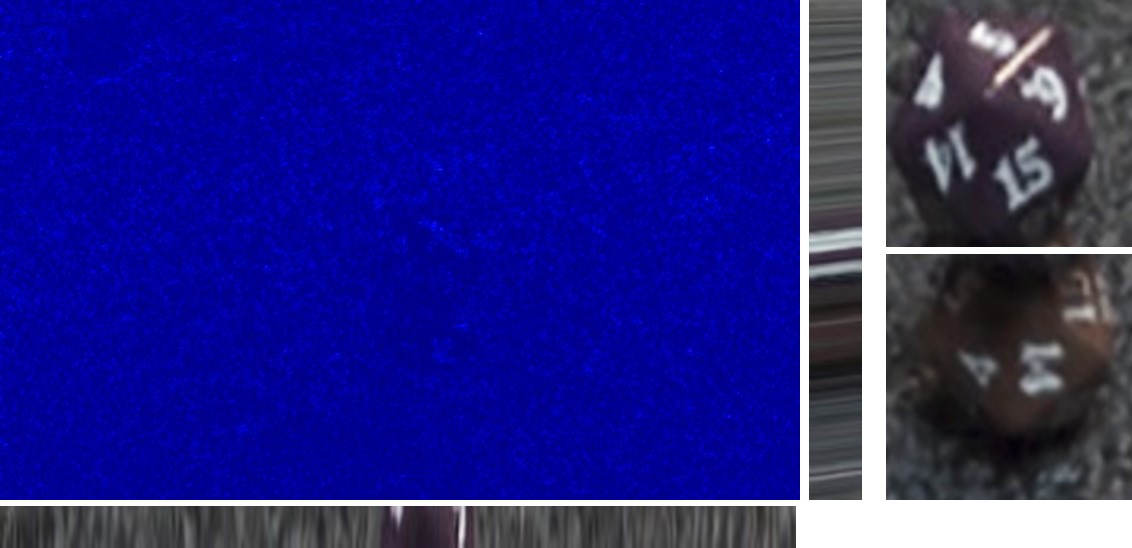}}{Proposed}~%

\end{center}
\caption{Qualitative evaluation on \textit{Flower} and \textit{Toys} dataset. EPIs shown are scaled and cropped for better visibility. The error map and sliced EPIs from our method indicate geometrically better light field across all datasets. From the top to bottom are Flower~\#5, Flower~\#15, Flower~\#94, Toys~\#29, and Toys~\#35. }
    \label{fig:Main_Compare}
\end{figure*}
\begin{figure*}[t]
\begin{center}

	{\includegraphics[width=0.25\linewidth]{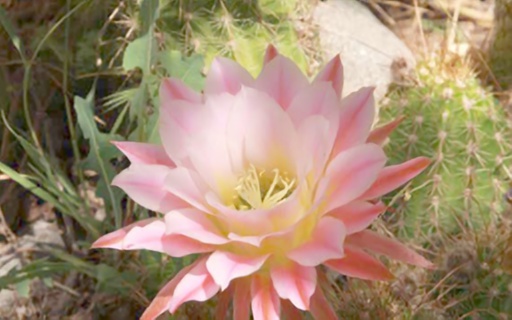}}~%
	{\includegraphics[width=0.25\linewidth]{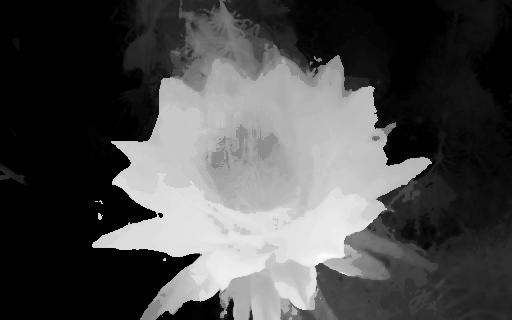}}~%
	{\includegraphics[width=0.25\linewidth]{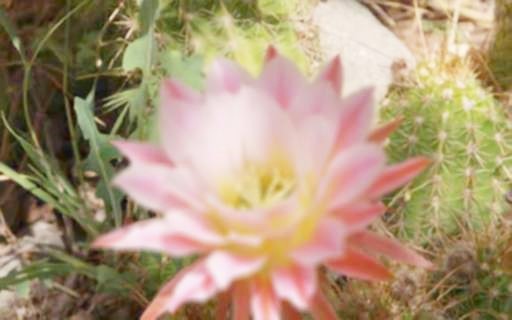}}~%
	{\includegraphics[width=0.25\linewidth]{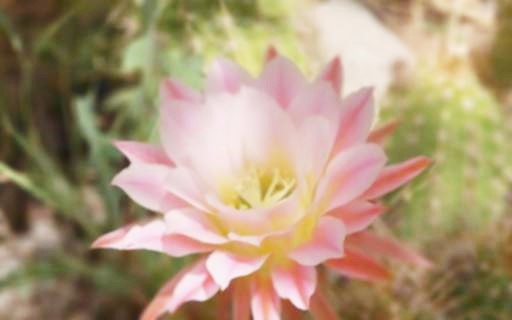}}~%
	
	\vspace{1.5mm}
	{\includegraphics[width=0.25\linewidth]{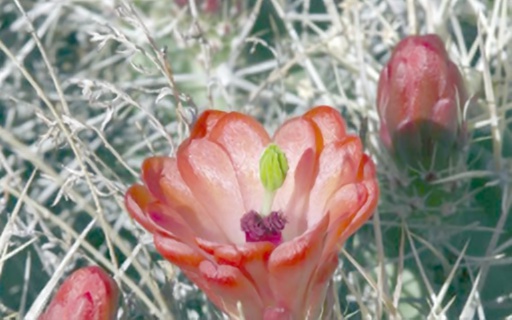}}~%
	{\includegraphics[width=0.25\linewidth]{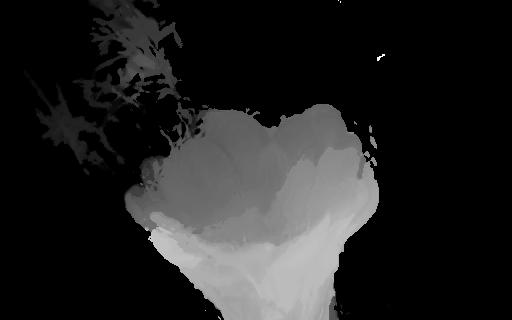}}~%
	{\includegraphics[width=0.25\linewidth]{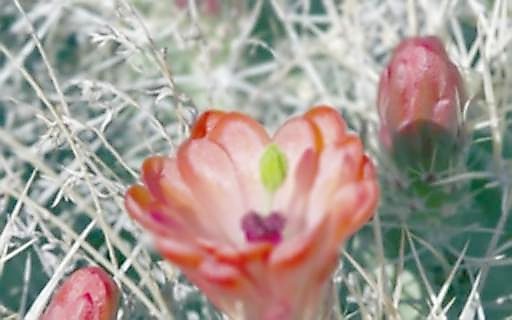}}~%
	{\includegraphics[width=0.25\linewidth]{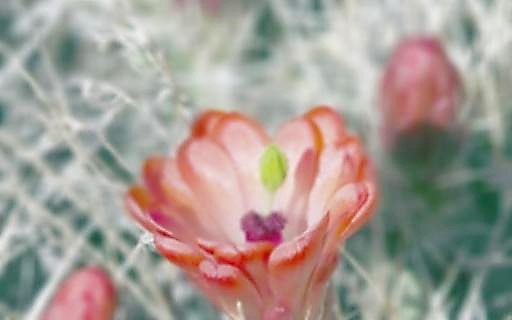}}~%
	
	\vspace{1.5mm}
	{\includegraphics[width=0.25\linewidth]{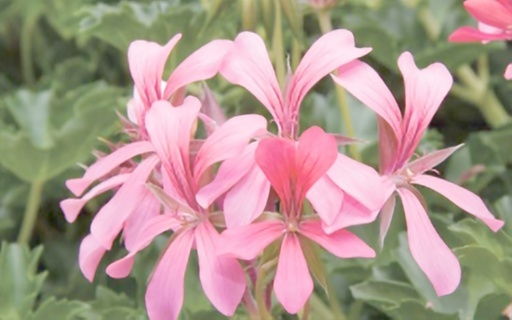}}~%
	{\includegraphics[width=0.25\linewidth]{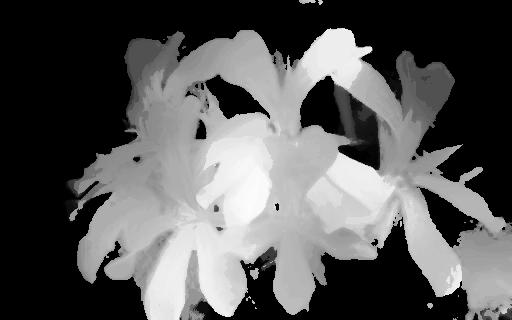}}~%
	{\includegraphics[width=0.25\linewidth]{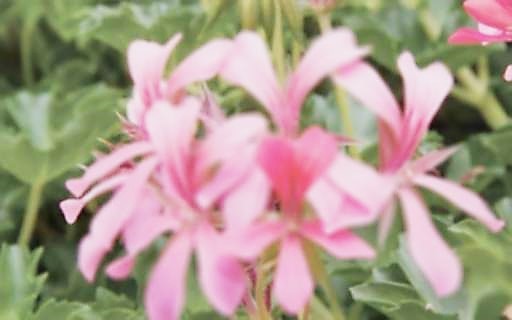}}~%
	{\includegraphics[width=0.25\linewidth]{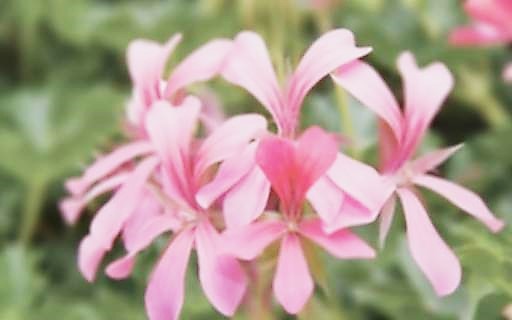}}~%
	
	\vspace{1.5mm}
	{\includegraphics[width=0.25\linewidth]{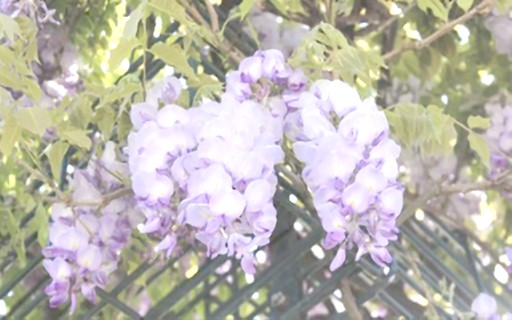}}~%
	{\includegraphics[width=0.25\linewidth]{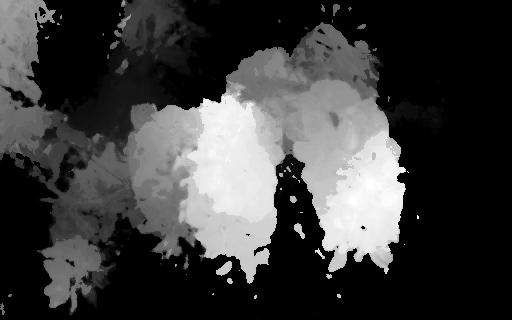}}~%
	{\includegraphics[width=0.25\linewidth]{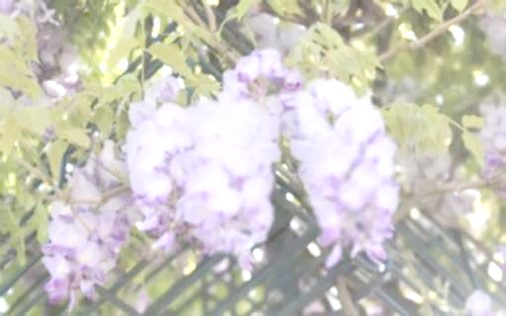}}~%
	{\includegraphics[width=0.25\linewidth]{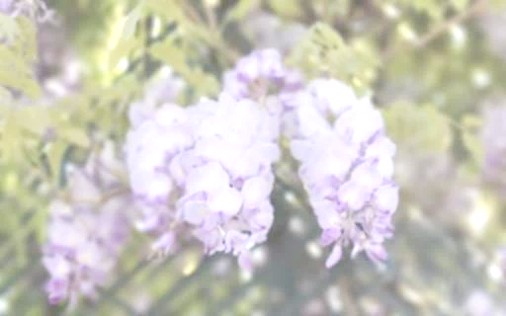}}~%

	\vspace{1.5mm}
    \stackunder[5pt]{\includegraphics[width=0.25\linewidth]{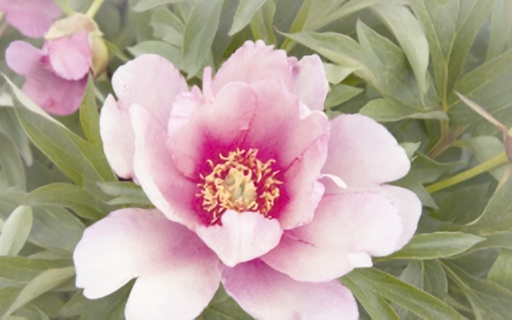}}{Input image}~%
	\stackunder[5pt]{\includegraphics[width=0.25\linewidth]{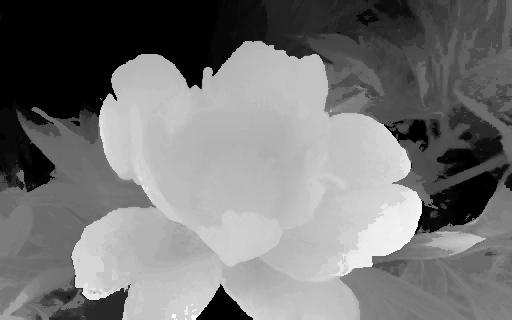}}{Estimated depth}~%
	\stackunder[5pt]{\includegraphics[width=0.25\linewidth]{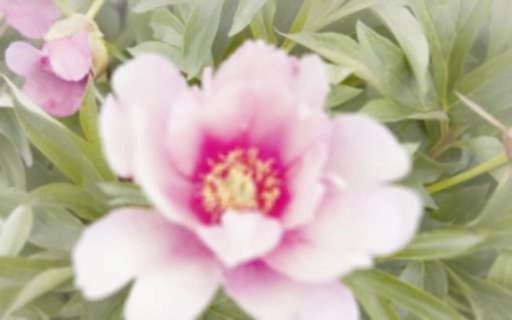}}{Background refocus}~%
	\stackunder[5pt]{\includegraphics[width=0.25\linewidth]{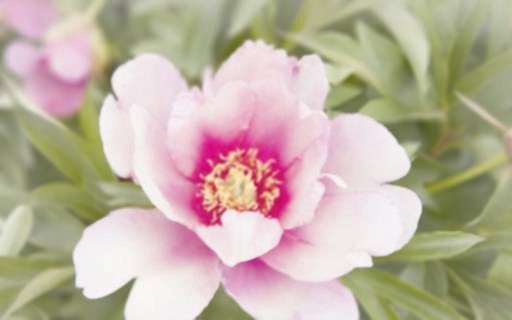}}{Foreground refocus}~%

\end{center}
\caption{Experiment on real single images and applications on the synthesized light field.}
    \label{fig:Single_Image}
\end{figure*}
\section{Experimental Results}
\label{sec:Experiment}
We evaluate the performance of the proposed framework both qualitatively and quantitatively.
We then compare its performance with the state-of-the-art method of light field angular synthesis from a single image ~\cite{Srinivasan_ICCV17}.

We utilize two datasets, {\em i.e.} \textit{Flower} and \textit{Toys}~\cite{Srinivasan_ICCV17}.
For the evaluation, all networks are re-trained with the corresponding dataset.
For evaluating Srinivasan~{\em et. al}~\cite{Srinivasan_ICCV17}, we use the author's original code.
Readers are recommended to view the supplementary video for better understanding and elaborated results.
\begin{table}[]
\centering
\resizebox{\linewidth}{!}{%
\begin{tabular}{@{}cccc@{}}
\toprule
Dataset         & Srinivasan    & Proposed LR   & Proposed HR   \\ \midrule
\textit{Flower} & 31.31 / 0.871 & 32.05 / \textbf{0.915} & \textbf{33.04} / 0.904 \\
\textit{Toys}   & 32.21 / 0.798 & 35.91 / \textbf{0.901} & \textbf{36.92} / 0.898 \\ \bottomrule
\end{tabular}
}
\caption{Average PSNR (in dB) and SSIM from \textit{Flower} (100 images) and \textit{Toys} (100 images) test set.}
  \label{table:Quantitative_General}
\end{table}
\subsection{Experiment on Available Light Field Dataset}
Light field images in the \textit{Flower} dataset mostly have a clear distinction between the background and foreground.
The flowers have a dominant color and are located in the foreground.
Unlike the \textit{Flower} dataset, \textit{Toys} dataset has more variance in object shape, color, and location in the image.
We utilize \textit{Toys} dataset to verify the performance of the proposed model in more general scenes.
\textit{Toys} dataset is significantly more complex and closer to the real world image.
We evaluate both the initial light field (low resolution) and the high resolution light field in PSNR and SSIM, which is shown in Table~\ref{table:Quantitative_General}.
In performing quantitative evaluation, we exclude the input image which is directly used as the center of the synthesized light field.
This is how numbers for the quantitative evaluation is obtained.
For each evaluation the network is trained on the corresponding dataset.
Note that we solve both super-resolution problems simultaneously and therefore our network need to solve more difficult problem than~\cite{Srinivasan_ICCV17}.
Nevertheless, the proposed method outperforms the state-of-the-art method on both low resolution (LR) and high resolution (HR) light field image.
The proposed network outperforms~\cite{Srinivasan_ICCV17}, which proves its capability for handling more complex and general scenes.

SSIM drop from LR to HR light field image stems from the second residual addition which causes slight blur in the erroneous region.
The spatial decoder tries to smooth out the erroneous region in the initial light field.
Figure~\ref{fig:Main_Compare} shows a qualitative comparison with the state-of-the-art work~\cite{Srinivasan_ICCV17}.
While \cite{Srinivasan_ICCV17} achieves high PSNR, it suffers from unpleasing artifacts around the edge and occluded region.
In addition, \cite{Srinivasan_ICCV17} cannot handle the scenes with multiple objects robustly.
\cite{Srinivasan_ICCV17} assumes that any pixel with a similar intensity is located in a close place.
\cite{Srinivasan_ICCV17} fails on general scene due to its dependency on finding a single object with a dominant color, which leads to inaccurate depth estimation.
As shown in the error map in Figure~\ref{fig:Main_Compare}, pixels on the flower in the backside have much error.
On the other hand, the proposed method successfully synthesizes a proper light field image due to the power of the proposed light field based loss function.
The error map and image patches show the proposed method handles occlusion better and have less artifact.

Generally the result of \cite{Srinivasan_ICCV17} on \textit{Toys} dataset shows incorrect EPI slope.
The slope direction is either reversed or flat indicating that the shifting direction of the object is incorrect.
\cite{Srinivasan_ICCV17} has difficulty in determining or inferring objects position.
On the contrary, the proposed method successfully synthesizes a proper light field.
\begin{table}[]
\centering
\resizebox{1.0\linewidth}{!}
{%
\begin{tabular}{@{}ccccccl@{}}
\toprule
Metric & L1    & $L_{g}$ & $L_{l}$ & $L_{g}+L_{l}$ & $\mathcal{S}$\\ \midrule
PSNR   & 31.69 & 32.02  & 32.09 & 32.82 & 32.86   \\
SSIM   & 0.877 & 0.887  & 0.892 & 0.901 & 0.901   \\ \bottomrule
\end{tabular}
}
\caption{Quantitative evaluation of each loss function's effect to the network trained using \textit{Flower} dataset. Each row represents the framework trained using only the corresponding loss, except for $\mathcal{S}$ (full framework trained without the image shifting). $L_{g}+L_{l}$ can also be inferred as results without regularization.}
  \label{table:Loss_Effect}

\end{table}
\begin{table}[]
\centering
\resizebox{1.0\linewidth}{!}
{%
\begin{tabular}{@{}cccc@{}}
\toprule
Combination & First     & Second    & Metric   (PSNR/SSIM)               \\ \midrule
1.          & App. Flow & -         & 32.55 / 0.891          \\
2.          & Intensity & -         & 32.31 / 0.891          \\
3.          & App. Flow & App. Flow & 32.35 / 0.903          \\
4.          & Intensity & Intensity & 32.32 / 0.903          \\
5.          & Intensity & App. Flow & 33.01 / 0.904          \\
6.          & App. Flow & Intensity & \textbf{33.04 / 0.904} \\ \bottomrule
\end{tabular}
}
\caption{Quantitative evaluation of multi-stage (first and second) residual estimation on \textit{Flower} dataset. Appearance flow and intensity denotes ${L}_{rf}(\textbf{x},\textbf{u})$ and ${L}_{ri}(\textbf{x},\textbf{u})$, respectively.}
  \label{table:Residual_Ablation}
\end{table}
\begin{figure*}[t]
\begin{center}
	\subfloat[Synthetic refocus on background and foreground, respectively.]
	{	
	{\includegraphics[width=0.25\linewidth]{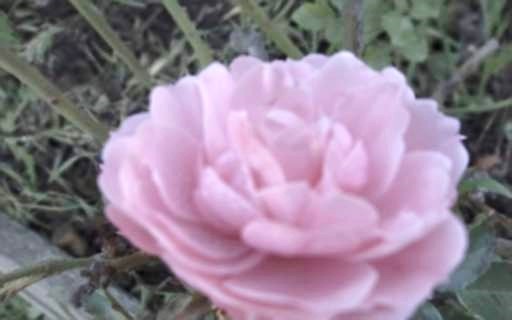}}~%
    {\includegraphics[width=0.25\linewidth]{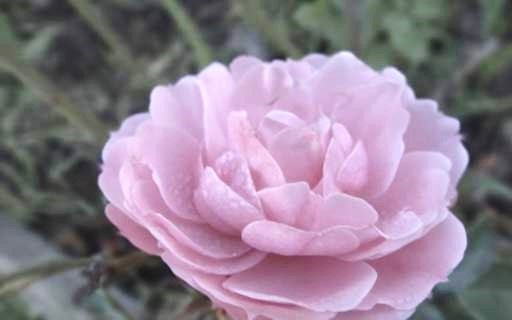}}~%
    {\includegraphics[width=0.25\linewidth]{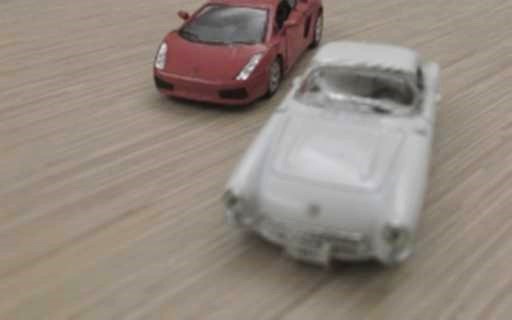}}~%
    {\includegraphics[width=0.25\linewidth]{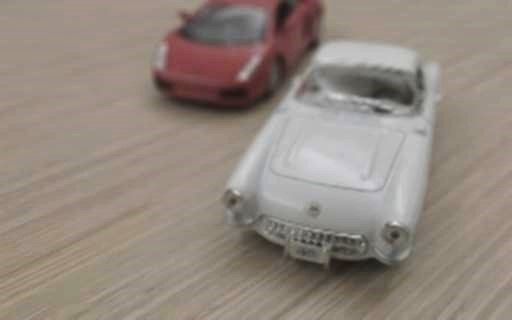}}
    }\\
	\subfloat[Estimated depth from the synthesized light field image.]
	{
	{\includegraphics[width=0.25\linewidth]{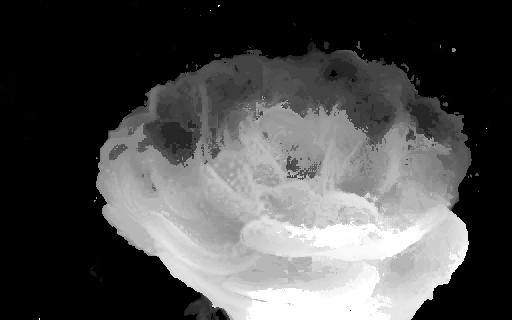}}~%
	{\includegraphics[width=0.25\linewidth]{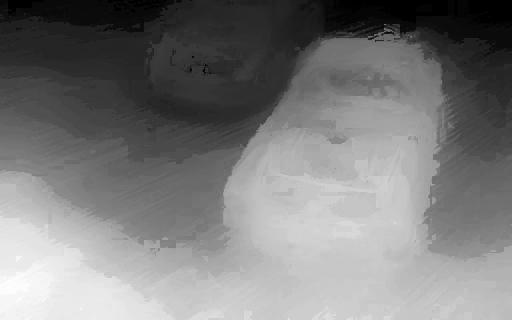}}~%
	}
	\subfloat[Estimated depth from the real light field image.]
	{
    {\includegraphics[width=0.25\linewidth]{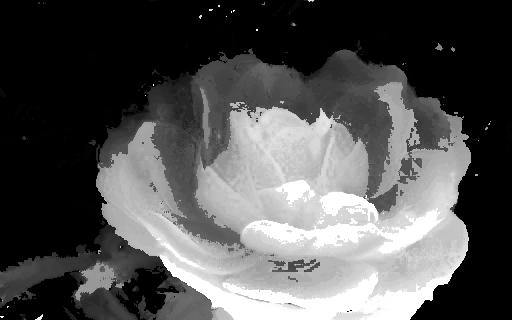}}~%
    {\includegraphics[width=0.25\linewidth]{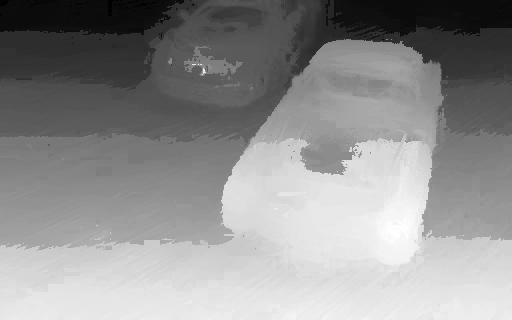}}~%
    }
\end{center}
\caption{Synthetic defocus result and depth estimation results using CAE~\cite{Williem_PAMI17} as a baseline.}
    \label{fig:Depth_Refocus_Result}
    \vspace{-4.0mm}
\end{figure*}
\begin{figure}[h]
\begin{center}
	{\includegraphics[width=0.33\linewidth]{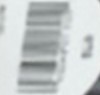}}~%
    {\includegraphics[width=0.33\linewidth]{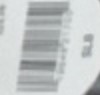}}~%
	{\includegraphics[width=0.33\linewidth]{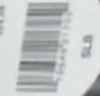}}%
    \vspace{3.0mm}

	{\includegraphics[width=0.33\linewidth]{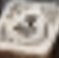}}~%
    {\includegraphics[width=0.33\linewidth]{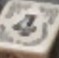}}~%
	{\includegraphics[width=0.33\linewidth]{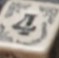}}%

	\subfloat[]{\includegraphics[width=0.33\linewidth]{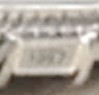}}~%
    \subfloat[]{\includegraphics[width=0.33\linewidth]{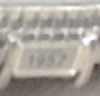}}~%
	\subfloat[]{\includegraphics[width=0.33\linewidth]{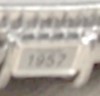}}%
\end{center}
\caption{Patches of spatial SR results. (a) LR input (256$\times$160). (b) First inference (512$\times$320). (c) Second inference (1024$\times$640).}
    \label{fig:Spatial_SR}
    \vspace{-4.0mm}
\end{figure}
\subsection{Ablation Study}
To reveal the discriminate power of the proposed light field synthesis network, we evaluate the performance of different loss functions, {\em i.e.} pixel-wise L1 loss, global loss, local loss, and local--global loss.
Table~\ref{table:Loss_Effect} shows the quantitative evaluation of each loss function's effect on the network.
The simple pixel-wise loss is outperformed by the proposed loss functions.
Meanwhile, local loss achieves satisfactory performance, and the combination with global loss leads to the best result.
Global and local losses complement each other by forcing local and global consistency in synthesizing SAIs.
Image shifting and regularization are proven to improve network performance as shown in $L_{g}+L_{l}$ and $\mathcal{S}$, respectively.

Additionally, we also show that the synthesized light field can be used for common light field applications, such as depth estimation and refocusing.
We test our network on real single image captured using a smartphone or downloaded from internet, as shown in Figure~\ref{fig:Single_Image}.
This shows our network capability to handle the images outside the training set distribution.
In addition, Figure~\ref{fig:Depth_Refocus_Result} shows convincing depth estimation and synthetic refocus.
The estimated depth in both figures shows the network can estimate both steep discontinuity or smooth discontinuity.
It confirm that the synthesized light field is geometrically correct and can be used in various applications of light field processing.

To justify our multi-stage residual estimation design, we perform an extensive experiment to identify the best combination.
Table~\ref{table:Residual_Ablation} shows that two-stage residual estimation produces the best light field quantitatively.
The order of residual addition does not cause any significant difference.
We hypothesize that at the first stage the network recovers the lost information during upsampling (anti-aliasing).
Thus, we observe strong gradient in the estimated residual and we believe this information strongly relates to the geometry information (appearance flow).
On the other hand, the second stage focuses on the homogeneous region. Therefore, it relates to the pixel color.
Note that we try to increase the multi-stage residual into three stages, which does not yield any meaningful boost to the performance.

We choose to perform angular resolution first as it saves the memory needed to train the network.
Upsampling the spatial resolution of the input image first will lead to large memory consumption in estimating the appearance flow.
The process of reconstructing the light field is expensive during training stage.
Performing angular upsampling first gives us extra information to help the spatial super resolution.
In addition, using residual approach it acts as a post-processing module while also solving the spatial SR problem.

Finally, we show the scalability of the proposed spatial SR module.
Since there is no available light field dataset in HR, we use similar technique by~\cite{Schoher_CVPR18} to predict image in HR.
To obtain HR light field we run the network twice sequentially.
During the second inference, we only take the estimated residual to enhance the previously estimated light field.
Since no HR ground truth is available, we show qualitatively in Figure~\ref{fig:Spatial_SR} that 4x as well as 2x spatial SR are performed successfully on the proposed framework.

\section{Proposed Network Structure}
In this section, we provide details of the proposed network structure that takes the luminance component of the input.
The dimension is represented as spatial (width and height) and channel or filter.
Each convolution has the bias initialized with 0 and the weight initialized using the He initializer~\cite{He_ICCV15}.
The image shifting operation is performed using \textit{tf.translate}.
Mean and variance for computing loss are obtained using \textit{tf.nn.moments}.
The order of light field stack is the column- or $\textbf{u}$ based.
The angular and spatial network details are tabulated on Table~\ref{tab:Network} and~\ref{tab:Network2}, respectively.
Every convolution is followed by Leaky ReLU activation function.
Convolution with stride~1, convolution with stride~2, transpose convolution, and convolution without activation function are denoted as `Conv', `Conv*', `Conv$^t$', and `Conv$^o$'.
Concatenate operation on channel axis is denoted as `Concat'.
\begin{table*}[h]
\centering
\resizebox{0.75\textwidth}{!}{
\begin{tabular}{|c|c|c|c|c|c|}
\hline
Network    & Layer & Operation         & Filter Size & Input Size     & Output Size     \\ \hline
           & 1     & Conv + Conv       & (3, 3, 16)  & (128, 128, 1)  & (128, 128, 16)  \\ \cline{2-6}
           & 1     & Conv*             & (3, 3, 16)  & (128, 128, 16) & (64, 64, 16)    \\ \cline{2-6}
           & 2     & Conv + Conv       & (3, 3, 32)  & (64, 64, 16)   & (64, 64, 32)    \\ \cline{2-6}
           & 2     & Conv*             & (3, 3, 32)  & (64, 64, 32)   & (32, 32, 32)    \\ \cline{2-6}
Encoder    & 3     & Conv + Conv       & (3, 3, 64)  & (32, 32, 32)   & (32, 32, 64)    \\ \cline{2-6}
           & 3     & Conv*             & (3, 3, 64)  & (32, 32, 64)   & (16, 16, 64)    \\ \cline{2-6}
           & 4     & Conv + Conv       & (3, 3, 128) & (16, 16, 64)   & (16, 16, 128)   \\ \cline{2-6}
           & 4     & Conv*             & (3, 3, 128) & (16, 16, 128)  & (8, 8, 128)     \\ \cline{2-6}
           & 5     & Conv + Conv       & (3, 3, 256) & (8, 8, 128)    & (8, 8, 256)     \\ \cline{2-6}
           & 5     & Conv*             & (3, 3, 256) & (8, 8, 256)    & (4, 4, 256)     \\ \hline
Bottleneck & 6     & Conv + Conv       & (3, 3, 512) & (4, 4, 256)    & (4, 4, 512)     \\ \hline
           & 7     & Conv$^t$ + Concat & (3, 3, 256) & (4, 4, 512)    & (8, 8, 256)     \\ \cline{2-6}
           & 7     & Conv + Conv       & (3, 3, 512) & (8, 8, 256)    & (8, 8, 256)     \\ \cline{2-6}
           & 8     & Conv$^t$ + Concat & (3, 3, 128) & (8, 8, 256)    & (16, 16, 128)   \\ \cline{2-6}
           & 8     & Conv + Conv       & (3, 3, 128) & (16, 16, 128)  & (16, 16, 128)   \\ \cline{2-6}
           & 9     & Conv$^t$ + Concat & (3, 3, 64)  & (16, 16, 128)  & (32, 32, 64)    \\ \cline{2-6}
Angular    & 9     & Conv + Conv       & (3, 3, 64)  & (32, 32, 64)   & (32, 32, 64)    \\ \cline{2-6}
Decoder    & 10    & Conv$^t$ + Concat & (3, 3, 64)  & (32, 32, 64)   & (64, 64, 64)    \\ \cline{2-6}
           & 10    & Conv + Conv       & (3, 3, 64)  & (64, 64, 64)   & (64, 64, 64)    \\ \cline{2-6}
           & 11    & Conv$^t$ + Concat & (3, 3, 64)  & (64, 64, 64)   & (128, 128, 64)  \\ \cline{2-6}
           & 11    & Conv + Conv       & (3, 3, 64)  & (128, 128, 64) & (128, 128, 64)  \\ \cline{2-6}
           & 12    & Conv + Conv       & (3, 3, 64)  & (128, 128, 64) & (128, 128, 64)  \\ \cline{2-6}
           & 12    & Conv$^o$          & (3, 3, 128) & (128, 128, 64) & (128, 128, 128) \\ \hline
\end{tabular}
}
\caption{The proposed angular decoder structure.
}
\label{tab:Network}
\end{table*}

\begin{table*}[h]
\centering
\resizebox{0.75\textwidth}{!}{
\begin{tabular}{|c|c|c|c|c|c|}
\hline
Network   & Layer & Operation         & Filter Size & Input Size      & Output Size     \\ \hline
          & 7     & Conv$^t$          & (3, 3, 256) & (4, 4, 512)     & (8, 8, 256)     \\ \cline{2-6}
          & 7     & Conv + Conv       & (3, 3, 256) & (8, 8, 256)     & (8, 8, 256)     \\ \cline{2-6}
          & 8     & Conv$^t$          & (3, 3, 128) & (8, 8, 256)     & (16, 16, 128)   \\ \cline{2-6}
          & 8     & Conv + Conv       & (3, 3, 128) & (16, 16, 128)   & (16, 16, 128)   \\ \cline{2-6}
          & 9     & Conv$^t$          & (3, 3, 64)  & (16, 16, 128)   & (32, 32, 64)    \\ \cline{2-6}
          & 9     & Conv + Conv       & (3, 3, 64)  & (32, 32, 64)    & (32, 32, 64)    \\ \cline{2-6}
          & 10    & Conv$^t$          & (3, 3, 64)  & (32, 32, 64)    & (64, 64, 64)    \\ \cline{2-6}
          & 11    & Conv$^t$ + Concat & (3, 3, 64)  & (64, 64, 64)    & (128, 128, 64)  \\ \hline
Residual  & 11    & Conv + Conv       & (3, 3, 64)  & (128, 128, 64)  & (128, 128, 64)  \\ \cline{2-6}
Flow      & 12    & Conv$^t$ + Concat & (3, 3, 192) & (128, 128, 64)  & (256, 256, 192) \\ \hline
          & 12    & Conv + Conv       & (3, 3, 192) & (256, 256, 192) & (256, 256, 192) \\ \cline{2-6}
          & 11*   & Conv$^t$ + Concat & (3, 3, 64)  & (64, 64, 64)    & (128, 128, 64)  \\ \hline
Residual  & 11*   & Conv + Conv       & (3, 3, 64)  & (128, 128, 64)  & (128, 128, 64)  \\ \cline{2-6}
Intensity & 12*   & Conv$^t$ + Concat & (3, 3, 192) & (128, 128, 64)  & (256, 256, 192) \\ \hline
          & 12*   & Conv + Conv       & (3, 3, 192) & (256, 256, 192) & (256, 256, 192) \\ \hline
\end{tabular}
}
\caption{The proposed spatial decoder structure. Layer 7 is the continuation of bottleneck.
}
\label{tab:Network2}
\end{table*}

\section{Training Details}
The proposed network is implemented using TensorFlow~\cite{Abadi_OSDI16}.
We crop the light field spatial resolution into random patches of 128$\times$128$\times$8$\times$8 for training.
Full light field resolution is used for inference.
The proposed network is trained in an end-to-end fashion for 280,000 iterations with a batch size of 1.
The angular decoder is trained alone for 100,000 iterations while the spatial decoder is freezed.
Hyperparameters ~$\eta$, $\lambda_l$, $\lambda_g$, $\lambda_{tv}$, and $\lambda_{sr}$ are set to 0.8, 1.0, 10, $1e^{-4}$, and 10.0, respectively.
We augment the input with random gamma ranging from 0.4 to 1.0.
The training input is also randomly crop around the center region.
The input augmentation is designed to avoid overfitting.
We use Adam optimizer~\cite{Kingma_ICLR15} as our optimization algorithm with the default parameters.
Training is performed for approximately 1 day on NVIDIA GTX 1080Ti GPU with 11GB of memory and Intel i7-7700 @3.60 GHz CPU with 16GB of memory.

\section{Conclusion}
In this paper, we proposed an end-to-end deep model for joint angular and spatial light field SR from a single image.
Novel light field based loss functions were introduced to preserve spatio-angular consistency and to remove the dependency of pixel intensity.
Joint end-to-end framework were presented to solve both problem simultaneously.
The experimental results showed that the proposed network outperformed the state-of-the-art algorithm qualitatively and quantitatively.
In addition, the proposed method can also be generalized to various scene, such as \textit{toys} and internet images.
Future work includes wide-baseline light field synthesis, and higher degree of spatial SR.

\begin{IEEEbiography}[{\includegraphics[width=1in,height=1.25in,clip,keepaspectratio]{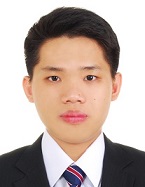}}]{Andre Ivan} (S'18-M'19) received his B.CompSc degree in computer science from Bina Nusantara University, Indonesia, in August 2017. He received his M.S. degree from Inha University, Korea, in August 2019. He was a visiting student in Center for Visual Computing, University of California, San Diego (UCSD). He is currently working for Pretia Technology in Japan as a computer vision researcher and development engineer. His research interests include computer vision, computational photography, deep learning, and GPGPU. He is a student member of the IEEE.
\end{IEEEbiography}

\begin{IEEEbiography}[{\includegraphics[width=1in,height=1.25in,clip,keepaspectratio]{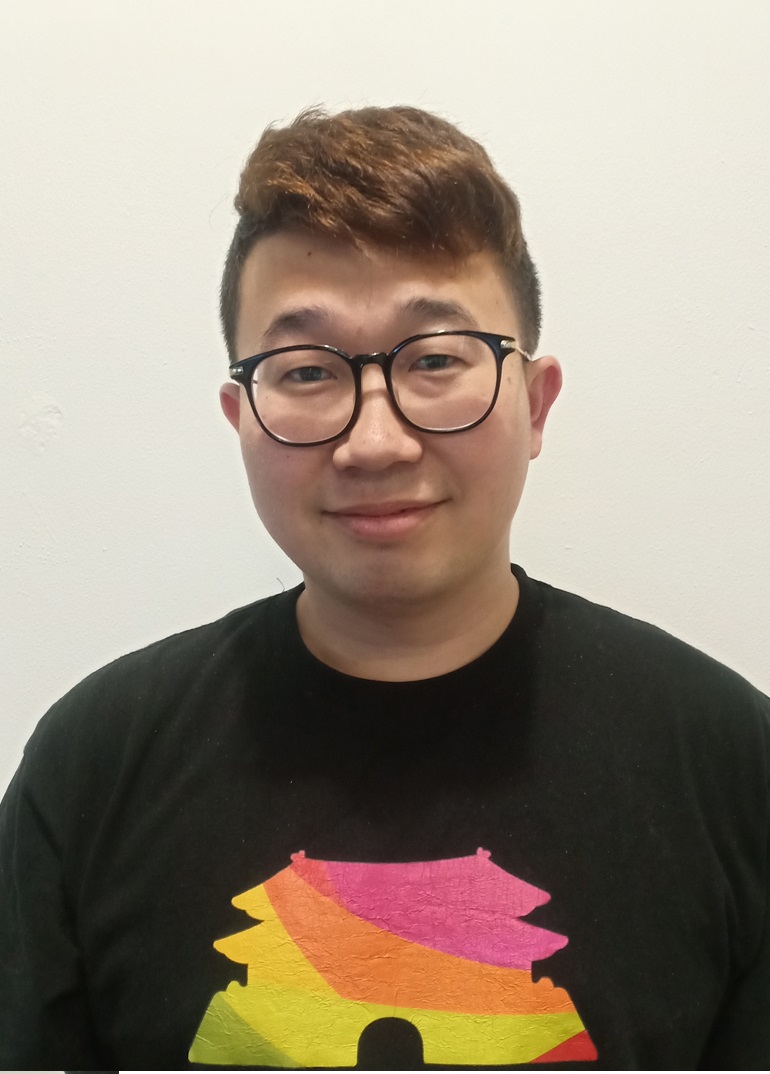}}]{Williem} (S'14-M'18) received his B.CompSc. degree in computer science from Bina Nusantara University, Indonesia, in 2011, and the Ph.D. degree in information and communication engineering from Inha University, Korea, in August 2017. From September 2017, he has been in School of Computer Science, Bina Nusantara University, Indonesia, as a faculty member. Currently, he is the chief executive officer of Verihubs since June, 2019. His research focuses on computer vision, computational photography, and image processing. Recent research topics are depth estimation of various conditions and light field applications. He is a member of the IEEE.
\end{IEEEbiography}

\begin{IEEEbiography}[{\includegraphics[width=1in,height=1.25in,clip,keepaspectratio]{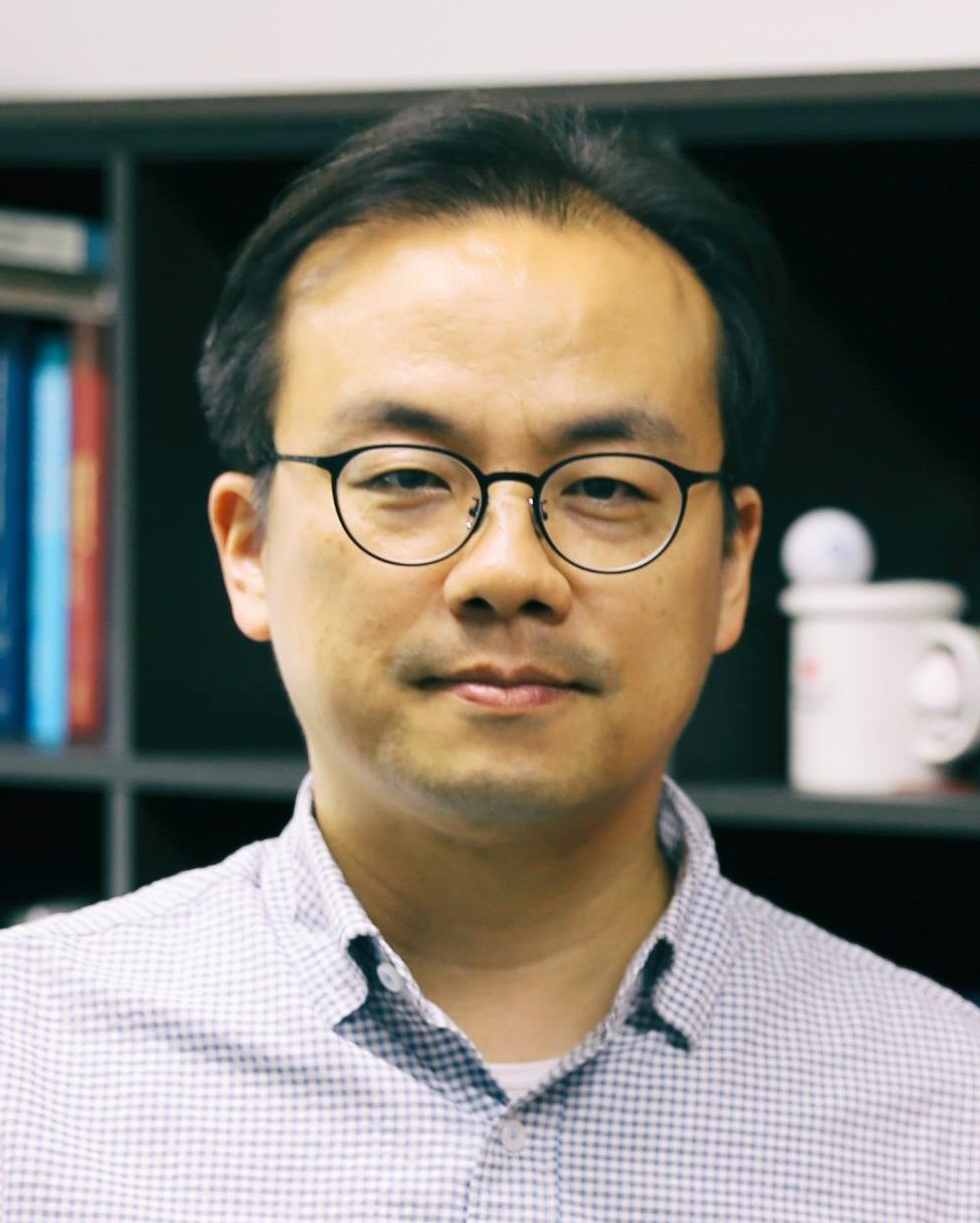}}]{In Kyu Park} (S'96-M'01-SM'14) received the B.S., M.S., and Ph.D. degrees from Seoul National University (SNU) in 1995, 1997, and 2001, respectively, all in electrical engineering and computer science. From September 2001 to March 2004, he was a Member of Technical Staff at Samsung Advanced Institute of Technology (SAIT). Since March 2004, he has been with the School of Information and Communication Engineering, Inha University, where he is a full professor. From January 2007 to February 2008, he was an exchange scholar at Mitsubishi Electric Research Laboratories (MERL). From September 2014 to August 2015, he was a visiting associate professor at MIT Media Lab. From July 2018 to June 2019, he was a visiting scholar at University of California, San Diego. Dr. Park's research interests include the joint area of computer vision and graphics, including 3D shape reconstruction from multiple views, image-based rendering, computational photography, deep learning, and GPGPU for image processing and computer vision. He is a senior member of IEEE and a member of ACM.
\end{IEEEbiography}

\EOD

\end{document}